# Increased Brightness and Reduced Efficiency Droop in Perovskite Quantum Dot Light-Emitting Diodes using Carbazole-Based Phosphonic Acid Interface Modifiers

Gillian Shen[1], Yadong Zhang[2], Julisa Juarez[1], Hannah Contreras[1], Collin Sindt[3], Yiman Xu[4], Jessica Kline[1], Stephen Barlow[2], Elsa Reichmanis[4], Seth R. Marder[2,3,5], David S. Ginger[1]

[1]Department of Chemistry, University of Washington, Seattle, WA 98195, United States
[2]Renewable & Sustainable Energy Institute, University of Colorado Boulder, Boulder, CO 80309, United States
[3]Department of Chemical and Biological Engineering, University of Colorado Boulder, Boulder, CO 80309, United States
[4]Department of Chemical and Biomolecular Engineering, Lehigh University, Bethlehem, PA 18015, United States
[5]Department of Chemistry, University of Colorado Boulder, Boulder, CO 80309, United States

**Abstract**

We demonstrate the use of [2-(9H-carbazol-9-yl)ethyl]phosphonic acid (2PACz) and [2-(3,6-di-*tert*-butyl-9H-carbazol-9-yl)ethyl]phosphonic acid (t-Bu-2PACz) as anode modification layers in metal-halide perovskite quantum dot light-emitting diodes (QLEDs). Compared to conventional QLED structures with PEDOT:PSS (poly(3,4-ethylenedioxythiophene) polystyrene sulfonate)/PVK (poly(9-vinylcarbazole)) hole-transport layers, QLEDs made with phosphonic acid (PA)-modified indium tin oxide (ITO) anodes show an over 7-fold increase in brightness, achieving a brightness of 373,000 cd m$^{-2}$, one of the highest brightnesses reported to date for colloidal perovskite QLEDs. Importantly, the onset of efficiency roll-off, or efficiency droop, occurs at ~1000-fold higher current density for QLEDs made with PA-modified anodes compared to control QLEDs made with conventional PEDOT:PSS/PVK hole transport layers, allowing the devices to sustain significantly higher levels of external quantum efficiency at a brightness of >10$^5$ cd m$^{-2}$. Steady-state and time-resolved photoluminescence measurements indicate these improvements are due to a combination of multiple factors, including reducing quenching of photoluminescence at the PEDOT:PSS interface and reducing photoluminescence efficiency loss at high levels of current density.

## Introduction

Colloidal quantum dot semiconductors have shown remarkable promise in applications[1,2] ranging from photovoltaics,[3] light-emitting diodes,[4–8] and photodiodes,[2,9] to single photon emission for quantum information science.[10–14] Metal-halide perovskite quantum dots in particular have emerged as efficient emitters[15,16] with near-unity photoluminescence quantum yields,[17,18] broad bandgap tailorability via compositional tuning,[8,15,19,20] narrow emission linewidths of <22 nm,[1,21,22] and the potential for additive manufacturing from solution.[23,24] Metal-halide perovskite quantum dot light-emitting diodes (QLEDs) have shown promise as a platform for color pure and efficient light-emission.[4,8,25–31] Compared with bulk 3D metal-halide perovskites, colloidal quantum dots offer larger exciton binding energies,[23,32] near-unity photoluminescence quantum efficiency,[10,33] and improved resistance to halide segregation.[8] However, the brightness of electrically driven perovskite QLEDs has to date lagged behind that of LEDs based on bulk

3D and 2D metal-halide perovskites, with the brightest QLEDs based on colloidal perovskite quantum dots remaining below 100,000 cd m$^{-2}$ [4,5,27] while the brightness of bulk perovskite LEDs (PeLEDs) has exceeded 473,000 cd m$^{-2}$ via the use of in-situ formed core-shell nanostructures to achieve a charge confinement effect.[34] High brightness is a key parameter needed for next-generation near-eye display applications such as augmented reality (AR) and virtual reality (VR) technologies. [32]

Like II-VI and III-V QLEDs, the brightness of halide perovskite QLEDs is often limited by efficiency droop, wherein the external quantum efficiency (EQE) of a device falls as brightness increases.[35–37] Indeed, this droop can be particularly significant for perovskite QLEDs, which typically exhibit peak EQEs at ~0.1 -1 mA cm$^{-2}$, beyond which EQE begins to decline.[4,27,38] Literature reports attribute efficiency droop to a combination of Joule heating,[35–37] Auger effects;[39] and charge injection imbalance.[37]

Herein we explore the use of dipolar phosphonic acid (PA) layers as interface modifiers in colloidal perovskite QLEDs and study their impact on brightness and efficiency droop. PA-based layers, often referred to as self-assembled monolayers (SAMs), have been widely studied in the context of organic electronics[40–46] and have recently been applied in metal-halide perovskite photovoltaics, [47–55] yet studies of their effects on the performance of perovskite LEDs have remained limited.[51,53] Here, we show that replacing the popular PEDOT:PSS/PVK injection layer with a PA-based interlayer leads to an increase in brightness and a significant suppression of efficiency roll-off (efficiency droop). To date, this is the first work, to our knowledge, to report on the effects of PA interfaces on hole-injection into perovskite QLEDs.

**Results**

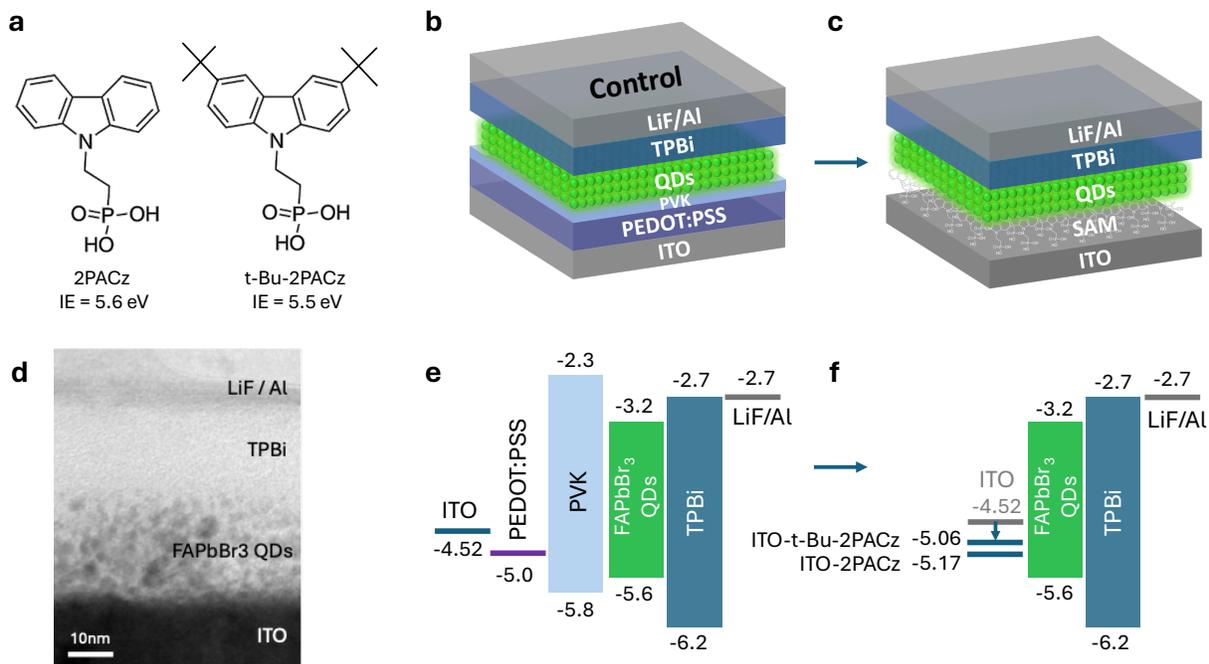

**Figure 1.** Phosphonic acid and device structures: a) molecular structures and estimated ionization energies of the phosphonic acids studied, 2PACz and t-Bu-2PACz respectively, b) schematic of control device structure, c) schematic of 2PACz and t-Bu-2PACz device structures, d) cross-sectional TEM image of phosphonic acid based device structure, e) energy-alignment diagram of the control devices, f) energy-

alignment diagram of the phosphonic acid devices with ITO, ITO-2PACz, and ITO-t-Bu-2PACz Fermi level energy with respect to vacuum, all reported in eV, measured by UPS (Figure S3) with energy levels of QDs, TPBi and LiF/Al derived from UPS and optical measurements by Chin et al..[56]

We compare colloidal perovskite QLEDs made with conventional hole transport layers (HTLs), PEDOT:PSS and PEDOT:PSS/PVK, where we modified the hole-injecting ITO interface with one of two phosphonic acids, [2-(9*H*-carbazol-9-yl)ethyl]phosphonic acid (2PACz)[46] or a related derivative, [2-(3,6-di-*tert*-butyl-9*H*-carbazol-9-yl)ethyl]phosphonic acid (t-Bu-2PACz), the structures of which are shown in Figure 1a. The molecules are synthesized following the schemes in Figures S1-S2. The PA moiety serves as an anchoring group that chemically binds to the surface sites of the indium tin oxide (ITO) in a bidentate or tridentate fashion, [40,52] enabling the formation of a stable and uniform monolayer or near-monolayer when deposited under optimal conditions. We deposited the PAs by a cosolvent strategy, which Liu et al. reported produces improved coverage vs. depositing directly from alcoholic solvents, and we verified the surface coverage using XPS (Figure S4).[50]

We prepared the quantum dot active layer material following an existing literature report, [56] with some modifications, using a room temperature, in-air ligand-assisted reprecipitation technique (LARP), described in detail in the Experimental Methods. In brief, we combined formamidinium bromide (FABr) with $PbBr_2$ in DMF to form the precursor solution with a 2:1 molar ratio of $FABr:PbBr_2$; we then added this solution dropwise into a vigorously stirred solution of toluene, butanol, *n*-octylamine, and oleic acid to form a cloudy dispersion which we centrifuged without an antisolvent for the collection of precipitate containing the desired QD product. We redissolved the QDs in toluene and centrifuged the dispersion a second time to collect the final clear supernatant. The second centrifugation aids in the removal of unreacted products, large particles, and residual impurities in the reaction flask. The resulting quantum dot dispersion has a solution photoluminescence quantum yield (PLQY) of 94% (for a dilute QD solution of OD~0.1 at 405 nm, excited at 405 nm with a monochromated Hg-Xe lamp at an intensity of ~3 $\mu Wcm^{-2}$, see Supporting Information).

We fabricated QLED devices using the synthesized quantum dots as an active layer, with control device structures of ITO/PEDOT:PSS(33 nm)/QDs(30 nm)/TPBi(40 nm)/LiF(1 nm)/Al(100 nm) and ITO/PEDOT:PSS(33 nm)/PVK(10 nm)/QDs(30 nm)/TPBi(40 nm)/LiF(1 nm)/Al(100 nm) in addition to the carbazole-phosphonic acid structures of ITO-2PACz/QDs(30 nm)/TPBi(40 nm)/LiF(1 nm)/Al(100 nm) and ITO-t-Bu-2PACz/QDs(30 nm)/TPBi(40 nm)/LiF(1 nm)/Al(100 nm). Figure 1b presents a schematic of the control device structure, while Figure 1c presents a schematic of the PA-modified structure. Figures 1e and 1f provide the energy alignment of the device layers in the control and PA-modified devices, respectively. Figure 1c shows the final cross-sectional profile of the PA-modified devices. In addition to simplifying the device structure and fabrication process, the use of the PAs also increases the work function of ITO (i.e., pushes the Fermi level deeper vs. vacuum), as determined by ultraviolet photoelectron spectroscopy (UPS) and shown in the schematic energy diagram of Figure 1f. Specifically, UPS indicates that the use of the PA layers shifts the work function of ITO from 4.52 eV to 5.06 eV in the case of t-Bu-2PACz, and to 5.17 eV in the case of 2PACz (see Supporting Information Figure S3). Compared to unmodified ITO (4.52 eV) the Fermi levels of the modified ITO substrates are thus much closer to the ionization energy corresponding to the valence band edge of the $FAPbBr_3$ QDs (5.6 eV, also measured by UPS).

Cross-sectional transmission electron microscopy (TEM) imaging confirms that the average thickness of the quantum dot layer is ~30 nm (Figure 1c). There is some overlap between the 40 nm thick evaporated TPBi and the quantum dot layer due to interstitial filling of the spaces between the nanocrystals. The nanocrystals exhibit a high degree of polydispersity and variations in interparticle spacing, which are visible in cross-sectional TEM images of the device and quantum dot layer at various levels of magnification (Figure 1c, Figure S5). TEM images of the QDs cast on a grid directly from solution also show some evidence of variations in interparticle spacing (Figure S6). The QD emission linewidth remains narrow, with a full width half maximum (FWHM) of 21 nm (Figure 2a), consistent with previous studies of this composition in QLED structures.[1,27,57] Figure S7 shows the Commission Internationale de l'Éclairage (CIE) coordinates for the devices with PEDOT:PSS/PVK, 2PACz, and t-Bu-2PACz interfaces, falling close to the REC2020 standard for green emission.

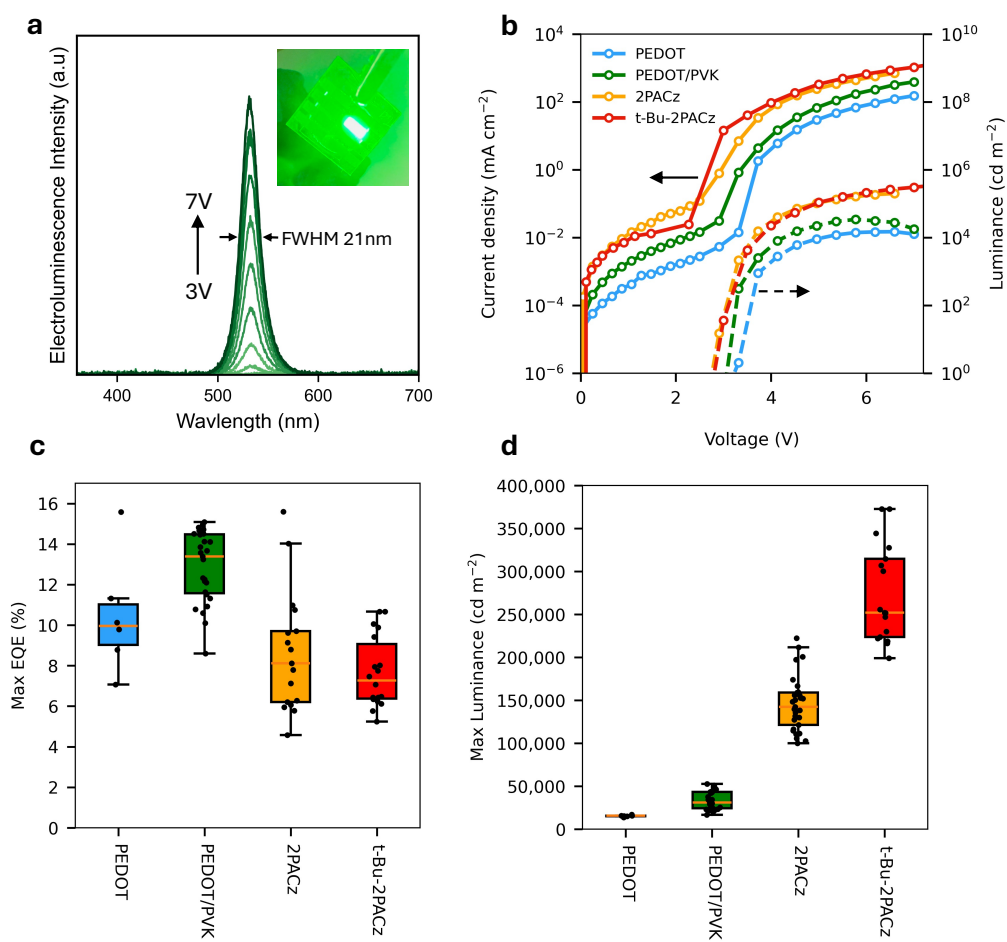

**Figure 2.** Device characteristics and statistics of 2PACz and t-Bu-2PACz QLED devices compared with control devices with PEDOT:PSS and PEDOT:PSS/PVK hole transport layers: a) electroluminescence spectra of the 2PACz devices at increasing driving voltages (inset) photograph of device under operation, b) current-voltage-luminance characteristics of the devices, c) EQE statistics, and d) maximum luminance statistics for N=6 PEDOT:PSS devices, N=28 PEDOT:PSS/PVK devices, N=29 2PACz devices, and N=18 t-Bu-2PACz devices showing significantly enhanced luminance in the PA-based devices.

We measured the current density-voltage-luminance (JVL) characteristics of the 2PACz and t-Bu-2PACz devices against the PEDOT:PSS and PEDOT:PSS/PVK control devices. Figure 2b presents the current-voltage-luminance curves, indicating that the addition of a PVK hole transport layer between the PEDOT:PSS and the quantum dots increases current density relative to the PEDOT:PSS-only device with quantum dots, consistent with previous reports.[37] The current density increases further when the control PEDOT:PSS/PVK transport layers are replaced with a 2PACz or t-Bu-2PACz monolayer on ITO. In conjunction with the higher current levels, we also observe significantly enhanced luminance in the PA devices – corresponding to a brightness enhancement of over 4-fold for the 2PACz and over 7-fold, reaching a brightness of 373,000 m$^{-2}$, for the t-Bu-2PACz devices – relative to the PEDOT:PSS/PVK control. The turn-on voltage, which we define as the knee point at which current begins to rapidly increase on the J-V curve, decreases from 3.3 V for PEDOT:PSS and 2.9 V for PEDOT:PSS/PVK to 2.5 V for 2PACz and 2.3 V for t-Bu-2PACz. The nominal energy-level diagram (Figure 1f) derived from UPS measurements and prior literature[56] suggests more facile injection should occur from PEDOT:PSS/PVK into the QDs than from the PA modified ITO. However, both PA devices show lower turn on voltages for current compared to the PEDOT:PSS/PVK. We attribute this discrepancy to the current being limited by barriers other than the HTL/QD interface in the PEDOT:PSS/PVK devices, likely the hole injection barrier at the interface between PEDOT:PSS and PVK, and/or barrier to electron extraction through the PVK interface. Moreover, it should also be borne in mind that the simple energy-level diagram shown ignores the effects of interfacial electron transfer and band bending.

Figures 2c and 2d show a summary of EQE and luminance characteristics of the 2PACz and t-Bu-2PACz devices as compared to control devices made with PEDOT:PSS only and combined PEDOT:PSS/PVK HTLs. The best devices made with PEDOT:PSS HTLs show a maximum EQE of 15.6% and a maximum luminance of 17,000 cd m$^{-2}$ while the best devices made with PEDOT:PSS/PVK HTLs have a maximum EQE of 15.0% and a maximum luminance of 52,600 cd m$^{-2}$. In contrast, the champion 2PACz device has a maximum EQE of 15.6% and a higher maximum luminance of 222,000 cdm$^{-2}$, showing over a 4-fold enhancement in luminance relative to the PEDOT:PSS/PVK controls. Finally, champion t-Bu-2PACz device has a maximum EQE of 10.7% and the highest maximum luminance of 373,000 cd m$^{-2}$, exhibiting over 7-fold luminance enhancement in the brightest devices relative to the PEDOT:PSS/PVK control. Table 1 summarizes the averages, standard deviations, and further statistics for luminance and EQE values across the conditions. The average luminance across all devices shows the same dependence on the selection of hole transport layer: PA-based devices consistently show higher luminances than either set of controls.

| Sample | Avg Luminance (cd m$^{-2}$) | Max Luminance (cd m$^{-2}$) | Avg EQE (%) | Max EQE (%) | Sample Count |
|---|---|---|---|---|---|
| PEDOT:PSS | 15,400 ± 1,030 | 17,000 | 10.4 ± 2.6 | 15.6 | 6 |
| PEDOT:PSS/PVK | 32,400 ± 11,600 | 52,600 | 12.7 ± 1.9 | 15.0 | 28 |
| 2PACz | 145,000 ± 31,700 | 222,000 | 8.6 ± 2.9 | 15.6 | 29 |
| t-Bu-2PACz | 274,000 ± 55,300 | 373,000 | 7.7 ± 1.7 | 10.7 | 18 |

**Table 1.** EQE and luminance statistics for PEDOT:PSS, PEDOT:PSS/PVK, 2PACz, and t-Bu-2PACz devices with average, standard deviation, and maximum values for luminance and EQE.

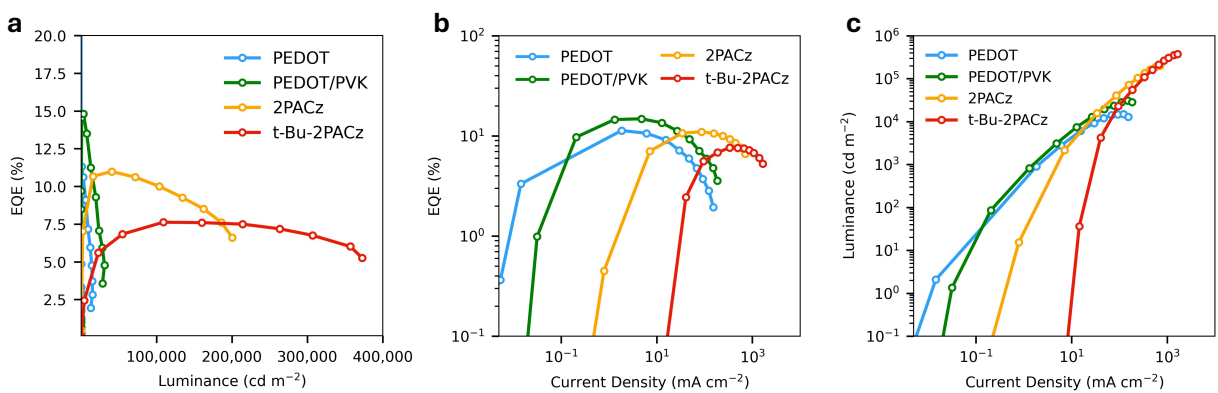

**Figure 3.** Further device characteristics of 2PACz and t-Bu-2PACz devices relative to controls: a) EQE-luminance characteristics, b) EQE-current density characteristics, c) and luminance-current density characteristics. The plots are parametric plots of voltage running from 0 V-7 V.

Further characterization of the devices under operation reveals other unique behaviors in devices with 2PACz and t-Bu-2PACz monolayers. Figure 3 presents EQE, luminance, and current density characteristics for representative devices made with each injection layer. The performance of the control devices is broadly consistent with reports of similar architectures in the literature,[27,58] while the performance of the devices with 2PACz and t-Bu-2PACz layers show significant improvements in all performance parameters. Figure 3a demonstrates that while EQE begins to fall off rapidly for the PEDOT:PSS/PVK controls as luminance increases past 5,000 cd m$^{-2}$, the EQE of 2PACz only begins to fall off after 50,000 cd m$^{-2}$, while the EQE of t-Bu-2PACz remains near its maximum at over 300,000 cd m$^{-2}$. Figure 3b presents a plot of EQE vs. current density, which reveals that they follow similar trends as the EQE vs. luminance. Notably, the 2PACz and t-Bu-2PACz devices support significantly higher levels of current before experiencing efficiency roll-off: for the representative devices shown in Figure 3b, the EQE peaks at 1.8 mA cm$^{-2}$ for PEDOT:PSS only devices, while peaking at 4.8 mA cm$^{-2}$ for PEDOT/PVK devices, peaking at a higher level of 86 mA cm$^{-2}$ for 2PACz device, and peaking at a higher level still of 335 mA cm$^{-2}$ for t-Bu-2PACz devices. Clearly, replacing the organic HTL with the PA modifiers significantly improves roll off. This ability to withstand high levels of current under continuous operation could have significant implications for applications in electrically pumped lasing where high drive currents are required.[59] Current density-luminance plots in Figure 3c reveal that while luminance begins to decline as current density reaches 123 mA cm$^{-2}$ for PEDOT and 150 mA cm$^{-2}$ for PEDOT/PVK devices, it only begins to plateau around 706 mA cm$^{-2}$ in the 2PACz devices and continues to rise beyond 1660 mA cm$^{-2}$ for the t-Bu-2PACz devices. We now turn to consider the reasons for the superior performance of the devices made with 2PACz and t-Bu-2PACz layers.

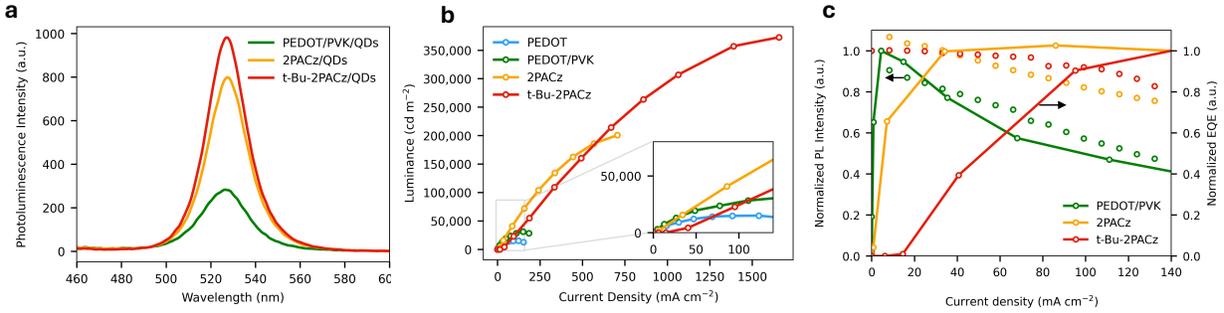

**Figure 4.** Photoluminescence and electroluminescence-based characterization of 2PACz and t-Bu-2PACz films and devices: a) steady-state photoluminescence measurements indicating QD films prepared on 2PACz- and t-Bu-2PACz-modified ITO exhibit enhanced photoluminescence intensities relative to QD films on an ITO/PEDOT:PSS/PVK control, b) luminance vs. device current density, showing brightness is enhanced in 2PACz devices and particularly in t-Bu-2PACz devices and that this is achieved at a higher current density, and c) normalized photoluminescence intensity and EQE vs. device current density, showing a suppressed loss of photoluminescence at high current injection densities in 2PACz and t-Bu-2PACz based samples correlating with the continued rise of EQE at high current injection densities.

First, steady state photoluminescence measurements (Figure 4a) indicate that PEDOT:PSS/PVK induces ~10-fold photoluminescence quenching relative to perovskite quantum dots on glass. Importantly, quantum dots on the 2PACz and t-Bu-2PACz-modified ITO exhibit a more moderate level of photoluminescence intensity quenching, with 2PACz showing a ~2.5-fold photoluminescence intensity enhancement and t-Bu-2PACz showing a ~3-fold photoluminescence enhancement relative to the PEDOT:PSS/PVK reference. Figure S8 shows time-resolved photoluminescence that exhibits similar trends. While the nominal band diagram in Figure 1e suggests that PVK should not quench the QD photoluminescence, we propose that either chemical effects from the acidic underlying PEDOT:PSS or tail states from the PVK valence band act to quench the photoluminescence of these perovskite QDs. This observation indicates that some of the improved brightness from the QLEDS made with 2PACz and t-Bu-2PACz layers comes from reduced photoluminescence quenching at the anode interface. However, this ~2-3 fold increase in photoluminescence intensity cannot account for the observed >7-fold increase in device brightness. More importantly, the change in static photoluminescence quenching alone would not account for the improvements in efficiency droop.

EQE depends on charge injection balance ($f_{balance}$), exciton formation probability ($f_{e-h}$), radiative-recombination efficiency ($\eta_{radiative}$), and outcoupling efficiency ($f_{outcoupling}$) and can be expressed as follows:
$$\mathrm{EQE} = f_{balance} \cdot f_{e-h} \cdot \eta_{radiative} \cdot f_{outcoupling}$$ [65]

To ascertain whether differences in film-forming properties on the different interfaces could lead to different light-outcoupling efficiencies, we performed darkfield scattering measurements (Figure S9) which showed no difference in darkfield scattering, suggesting differences in outcoupling ($f_{outcoupling}$), due to interfacial roughness, are minimal.

Bias-dependent carrier imbalance,[37] Joule heating,[35–37,60] and Auger recombination[61,62] (which increases the non-radiative decay rate at increasing current density) are commonly reported causes of efficiency droop.[37,39] To look for effects of non-radiative Auger recombination and Joule heating in the devices, we performed current density-dependent photoluminescence measurements on devices to observe correlations in electroluminescence and photoluminescence across current density levels in our devices.[39] Auger effects become increasingly pronounced at high carrier injection densities.[39,63,64] We used frequency-modulated laser excitation (405 nm, 28 mW cm$^{-2}$) at 1000 Hz coupled with lock-in detection to isolate the photoluminescence signal from a series of QLEDs under electroluminescence conditions under current densities from 0 to 140 mAcm$^{-2}$.[37,39] Figure 4c presents the lock-in-detected photoluminescence under various current densities: while all devices show photoluminescence quenching at increasing current density, the quenching is ~3-fold stronger in devices with PEDOT:PSS/PVK anodes than with t-Bu-2PACz. The control PEDOT:PSS/PVK devices show an absolute ~3.7% loss of photoluminescence intensity for every additional mA cm$^{-2}$ of current flowing, while the 2PACz devices show only ~2.4% photoluminescence loss for every additional mA cm$^{-2}$ and the t-Bu-2PACz devices show as little as ~1.2% photoluminescence loss for every additional mA cm$^{-2}$, indicating the clear suppression of Joule heating and/or Auger effects in the PA-based devices.

Figure 4c shows the close correlation between photoluminescence loss at high current densities ($\eta_{radiative}$) and a corresponding loss in EQE in the PEDOT:PSS/PVK devices. From this correlation, we conclude that a likely combination of Auger losses and Joule heating are responsible for the efficiency roll-off in the control PEDOT:PSS/PVK devices. In the 2PACz and t-Bu-2PACz devices, on the other hand, EQE continues to rise as photoluminescence loss occurs at a more moderate rate, indicating that charge injection balance is limiting EQEs at lower current levels, yet injection balance improves at higher current levels – ultimately leading to Auger and Joule heating effects only dominating beyond the $10^2$ to $10^3$ mA cm$^{-2}$ regime. Although carrier-injection balance is suboptimal at low voltages, it rises to its peak at high current levels, allowing the achievement of improved maximum brightness levels in the devices. To look for additional evidence of the role of Joule heating, we analyzed both the red tail of the LED electroluminescence[35,36,66–68] profile and its peak position[69] (Figures S10-S11). Although discrepancies exist in the absolute temperatures calculated using the two methods, both methods indicate that the control PEDOT:PSS/PVK devices rise to higher temperatures during device operation at lower drive currents and power levels, a product of more severe Joule heating effects at play in the devices. We speculate that this more severe heating could contribute to the early onset of roll-off effects at lower drive currents.[35,36,68]

From these data, we propose that the improved efficiency droop properties likely emanate from the reduction of photoluminescence quenching at high current densities (from a combination of reduced Auger recombination and improved thermal properties).

**Conclusions**
We study the anode interface modifiers, 2PACz and t-Bu-2PACz, in perovskite QLED structures, and show that they are effective in enhancing perovskite QLED brightness by suppressing efficiency droop (roll-off) owing to Auger and Joule heating effects. Injection current-dependent photoluminescence measurements indicate the presence of more severe photoluminescence quenching at higher injection current densities in the control PEDOT/PVK devices than the 2PACz or t-Bu-2PACz modified devices, correlating with the decline of EQE in the control devices providing evidence for the dominance of Joule heating and Auger effects in the roll-off behavior observed in the controls. The more moderate rates of photoluminescence decline at higher currents in the PA-modified devices enables EQEs to continue rising at high carrier injection densities. We believe that lower EQEs at lower injection densities in the 2PACz and t-Bu-2PACz devices are likely limited by current injection balance. The t-Bu-2PACz devices can operate at a high current density of >1 A/cm$^2$ and achieve a record brightness for perovskite QLEDs by suppression of Auger effects / Joule heating. These results suggest that continued improvements in interfacial engineering, either with phosphonic acid or polymers with properties engineered to inject into halide perovskite, are promising routes for improving the efficiency and stability of perovskite QLEDs.

## Experimental Methods

**Materials**
The following materials were obtained from the commercial sources stated: FABr (Greatcell Solar), PbBr$_2$ (TCI), DMF (Sigma Aldrich), oleic acid (Sigma Aldrich), octylamine (Sigma Aldrich), butanol (Sigma Aldrich), toluene (Sigma Aldrich), PVK (Sigma Aldrich), LiF (Sigma Aldrich), Merisuds (CH$_2$O), PEDOT:PSS (Ossila), TPBi (Ossila), Aluminum pellets (Kurt J. Lesker)

**General Synthetic Details**. $^1$H-NMR spectra, and $^{13}$C-NMR, $^{31}$P-NMR spectra were recorded with Bruker Avance 400 MHz spectrometer. ESI mass spectra were provided by the University of Colorado Boulder mass spectra facility laboratory. Elemental analyses were obtained from Atlantic Microlabs. (2-(9*H*-Carbazol-9-yl) ethanol (compound **3** in Figure S1)[70] was prepared in 79% yield from carbazole (**1**) and 2-bromoethanol (**2**) and converted to (2-(9*H*-carbazol-9-yl)-1-bromo-ethane (**4**)[71] according to literature methods. All other starting materials were purchased as reagent grade and used without further purification.

**Diethyl (2-(9*H*-carbazol-9-yl)ethyl)phosphonate.** A mixture of (2-(9*H*-carbazol-9-yl)-1-bromo-ethane (**4** in Figure S1, 4.0 g, 14.6 mmol) in triethyl phosphite (45 mL) was stirred for 20 h at 160 ºC (oil bath) under N$_2$. The excess amount of triethyl phosphite was removed under vacuum at 100 ºC to give diethyl (2-(9H-carbazol-9-yl)ethyl)phosphonate (**5** in Figure S1) as pale yellow liquid (4.8 g, 100%), which was used for the next step without future purification. $^1$H NMR,$^{13}$C{$^1$H} NMR and $^{31}$P{$^1$H} NMR spectra of diethyl [2-(9*H*-carbazol-9-yl)ethyl]phosphonate (**5**) in CDCl$_3$ are shown in Figure S14. $^1$H NMR (400 MHz, CDCl$_3$) $\delta$: 8.09 (dm, *J* = 8.0 Hz, 2H), 7.47 (td, *J* = 8.0 Hz, 2.0 Hz, 2H), 7.43 (dm, *J* = 8.0 Hz, 2.0 Hz, 2H), 7.25 (td, *J* = 8.0 Hz, 2.0 Hz, 2H), 4.62 (m, 2 H), 4.09 (m, 4H), 2.27 (m, 2H), 1.28 (td, *J* = 8.0 Hz, *J*$_{HP}$ = 1.6 Hz, 6H) ppm. $^{13}$C{$^1$H} NMR (100 MHz, CDCl$_3$) $\delta$: 139.63, 125.81, 123.07, 120.44, 119.23, 109.43, 61.87 (d, *J*$_{CP}$ = 6.0 Hz), 36.88, 25.29 (d, *J*$_{CP}$ = 137.0 Hz) 16.36(d, *J*$_{C-P}$ = 5.0 Hz) ppm. $^{31}$P{$^1$H} NMR (162 MHz, CDCl$_3$) $\delta$: 27.63 (s) ppm.

**[2-(9*H*-Carbazol-9-yl)ethyl]phosphonic acid (2PACz).** Bromotrimethylsilane (12.0 mL, 90.9 mmol) was added to diethyl (2-(9*H*-carbazol-9-yl)ethyl)phosphonate (**5** in Figure S1, 4.8 g, 14.5 mmol) in dry dichloromethane (80.0 mL) at room temperature under N$_2$. The reaction mixture was stirred at room temperature for 19 h. The solvent was removed under reduced pressure to give a colorless amorphous solid; methanol (30 mL) was added and stirred for 30 min and the water (100 mL) was added to the methanol solution. The mixture was then concentrated under reduced pressure to give **2PACz** as a white solid (4.0 g, 100%). $^1$H NMR,$^{13}$C{$^1$H} NMR and $^{31}$P{$^1$H} NMR spectra of **2PACz** in DMSO-*d*$_6$ are shown in Figure S15. $^1$H NMR (400 MHz, DMSO-*d*$_6$) $\delta$: 8.16 (d, *J* = 8.0 Hz, 2H), 7.55 (d, *J* = 8 Hz, 2H), 7.47 (td, *J* = 8.0 Hz, 2.0 Hz, 2H), 7.22 (td, *J* = 8.0 Hz, 2.0 Hz, 2H), 6.12 (s, br, 2H), 4.56 (m, 2H), 2.05 (m, 2H) ppm. $^{13}$C{$^1$H}

NMR (100 MHz, DMSO-$d_6$) δ: 139.86, 126.31, 122.74, 120.87, 119.41, 109.39, 37.84, 27.73 (d, $J_{CP}$ = 130 Hz) ppm. $^{31}$P{$^1$H} NMR (162 MHz, DMSO-$d_6$) δ: 21.58 (s) ppm. The above data are consistent with reported data in the literature.[72]

**Diethyl [2-(3,6-di-*tert*-butyl-9*H*-carbazol-9-yl)ethyl]phosphonate.** A mixture of 3,6-di-*tert*-butyl-9*H*-carbzole (**6** in Figure S2, 2.0 g, 7.16 mmol), diethyl (2-bromoethyl)phosphonate (**7**, 2.0 g, 8.16 mmol), $K_2CO_3$ (3.5 g, 25.32 mmol) in DMSO (7.0 mL) was stirred at 115 °C for 23 h. Water (100 mL) was added, the resulting brown solid was collected by filtration and washed with water. After drying, the crude product was purified by silica gel column chromatography using ethyl acetate as eluent. After removal of solvents under reduced pressure, the white solid was recrystallized from methanol/water to give pure product (compound **8** in Figure S2, 1.8 g, 56%). $^1$H NMR, $^{13}$C{$^1$H} NMR and $^{31}$P NMR spectra of diethyl [2-(3,6-di-*tert*-butyl-9*H*-carbazol-9-yl)ethyl]phosphonate (**8**) in CDCl$_3$ are shown in Figure S16. $^1$H NMR (400 MHz, CDCl$_3$,) δ: 8.12 (d, *J* = 2.0 Hz, 2H), 7.55 (dd, *J* = 8.0, 2.0 Hz, 2H), 7.36 (d, *J* = 8.0 Hz, 2H), 4.60 (m, 2H), 4.12 (m, 4H), 2.27 (m, 2H), 1.49 (s, 18H), 1.33 (t, *J* = 8.0 Hz, 6H) ppm. $^{13}$C{$^1$H} NMR (100 MHz, CDCl$_3$) δ: 142.10, 138.19, 123.49, 123.04, 116.45, 107.86, 61.87 (d, $J_{CP}$ = 7.0 Hz), 36.96, 34.70, 32.05, 25.43 (d, $J_{CP}$ = 136.0 Hz), 16.43 (d, $J_{CP}$ = 6.0 Hz) ppm. $^{31}$P{$^1$H} NMR (162 MHz, CDCl$_3$) δ: 27.86 (s) ppm. HRMS-ESI, Calcd for $C_{26}H_{39}NO_3P$ (MH$^+$): 444.2662; Found: 444.2658. Anal. Calcd for $C_{26}H_{38}NO_3P$: C, 70.40; H, 8.64; N, 3.16; Found: C, 70.67; H, 8.68; N, 3.52.

**[2-(3,6-Di-*tert*-butyl-9*H*-carbazol-9-yl)ethyl]phosphonic acid (t-Bu-2PACz).** Bromotrimethylsilane (2.6 g, 17 mmol) was added to a solution of diethyl [2-(3,6-di-*tert*-butyl-9*H*-carbazol-9-yl)ethyl]phosphonate (compound **8** in Figure S2, 1.5 g, 3.38 mmol) in dry dichloromethane (15 mL) under nitrogen at room temperature. The reaction mixture was stirred at room temperature for 22 h. The dichloromethane was removed under reduced pressure, methanol (20 mL) was added and the solution was stirred at room temperature for 30 min. Water (60 mL) was added into the methanol solution and the resulting white solid was collected by filtration and washed with water. After drying, the product was obtained as a white solid (1.3 g, 98%). $^1$H NMR, $^{13}$C{$^1$H} NMR and $^{31}$P{$^1$H} NMR spectra of **t-Bu-2PACz** in DMSO-$d_6$ are shown in Figure S17. $^1$H NMR (400 MHz, DMSO-$d_6$) δ: 8.20 (d, *J* = 2.0 Hz, 2H), 7.51 (dd, *J* = 8.0, 2.0 Hz, 2H), 7.41 (d, *J* = 8.0 Hz, 2H), 5.15 (s, br, 2H), 4.49 (m, 2H), 2.00 (m, 2H), 1.41 (s, 18H) ppm. $^{13}$C{$^1$H} NMR (100 MHz, DMSO-$d_6$) δ: 141.71, 138.41, 123.76, 122.75, 116.97, 108.72, 37.85, 34.89, 32.37, 27.92 (d, $J_{C-P}$ = 130.0 Hz) ppm. $^{31}$P{$^1$H} NMR (162 MHz, DMSO-$d_6$) δ: 21.71 (s) ppm. HRMS-ESI, Calcd for $C_{22}H_{31}NO_3P$ (MH$^+$): 388.2036; Found: 388.2032. Anal. Calcd for $C_{22}H_{30}NO_3P$: C, 68.20; H, 7.80; N, 3.62; Found: C, 68.47; H, 7.90; N, 3.57

**Synthesis of FAPbBr$_3$ Quantum Dots.** Quantum dots were synthesized following the preparation detailed by Chin et al., with small modifications.[56] The FABr precursor was prepared by dissolving 0.8 M FABr in DMF. The PbBr$_2$ precursor was prepared by dissolving 0.4 M PbBr$_2$ in DMF. 250 μL of each precursor was combined and vortexed to form 0.5 mL of precursor solution A. A separate solution B was prepared with 15 mL toluene, 6 mL butanol, 75 μL n-octylamine, and 900 μL oleic acid. For each synthesis, 450 μL of precursor A was added dropwise into solution B under vigorous stirring to form a cloudy dispersion of FAPbBr$_3$ NCs. The solution was centrifuged at 11,000rpm (~10,300 rcf) for 15 min; the supernatant was then discarded and the remaining precipitate was redispersed in toluene (3 mL). The redispersed solution was then centrifuged at ~4000 rpm (1,300 rcf) to form the supernatant of nanocrystal ink. This supernatant was filtered through a 0.2μm PTFE filter to form the final NC dispersion.

**Device Fabrication**. Pre-etched indium tin oxide (ITO) glass was sequentially sonicated in 2% Merisuds, deionized (DI) water, acetone, and isopropyl alcohol (IPA) for 20 min each, dried with a nitrogen gun, and vacuum-sealed before use. The substrates were UV-ozone plasma cleaned for 20 min before device fabrication (Harrick Plasma PDC-001-HP).

**PVK Control Devices**. PEDOT:PSS was spin coated at 4000 rpm for 40 s, followed by a 15 min anneal at 150 °C. The substrates were then transferred into a nitrogen-filled glovebox for subsequent layers. PVK (5 mg/mL) was spin coated at 4000 rpm for 50 s, followed by a 15 min anneal at 120 °C.

**2PACz and t-Bu-2PACz Devices**. 2PACz and t-Butyl 2PACz stock solutions were prepared by dissolving each in DMF at a concentration of 50 mg/mL, then diluted to a concentration of 1mg/mL in IPA by dissolving 20 μL of the stock in 980 μL of IPA. This solution was then spun coated onto the glass/ITO substrates at 3000 rpm for 30 s, with a 600rpm/s acceleration, then annealed for 10 min at 100 °C.

130 μL of the quantum dot solution was then dropped onto the 1.5 cm × 1.5 cm substrate and left to sit for 5 min before spinning at 1000 rpm for 60 s, followed by a 10 min anneal at 70 °C. The devices were then loaded into a thermal evaporation chamber (Angstrom engineering MB200B) and pumped down to $7\times10^{-7}$ Torr for the deposition of subsequent layers. 40 nm of TPBi was thermally evaporated without a shadow mask, followed by 1 nm of LiF and 100 nm of Al under a shadow mask.

**Device Characterization**. The LED EQE measurement was performed using a homebuilt setup with a large area silicon photodiode (Hamamatsu S3204-08), with a machined sample holder keeping the LED emitter surface exactly 1 mm from the photodetector.[73] The complete setup was housed in a dark enclosure. A geometry calculation was performed for the specific geometry and position of each pixel, with the assumption of a Lambertian emission profile, to find that the detector captures 90.1-97.2% of forward-emitted light depending on the exact pixel.[73] The sample holder was designed to block waveguided emission from the detector. A Keithley 2400 source meter unit is used to drive the device at bias voltage increments while measuring device current density, while a second Keithley 2400 source meter captures the photocurrent generated by the Si photodiode. EQE and luminance calculations were performed according to best practices recommendations from Anaya et al.[74] while accounting for the calibrated responsivity of the photodetector. The responsivity function of the photodetector is measured using an Oriel EQE setup with a 300 W Xe arc lamp light source and monochromator. Device emission spectra were measured using an Ocean Optics USB2000 spectrometer. The sensitivity of the spectrometer was calibrated using a HL-2P-INT-CAL white light calibration source, using a calibration file provided from the manufacturer.

**QD, Film, and Further Device Characterization**. Photoluminescence measurements were performed on a PerkinElmer Fluorescence Spectrometer LS 55. UV-Vis measurements were performed on a PerkinElmer UV/VIS/NIR Spectrometer Lambda 950. PLQY measurements were performed with a Hamamatsu Xenon/Mercury-Xenon Lamp as an excitation source, with emission measured with a Hamamatsu Model A10104-01 integrating sphere coupled to a Hamamatsu Photonic Multi-Channel Analyzer C10027. PLQY Calibrations were performed with Rhodamine 6G as a reference. TEM measurements were taken on the Tecnai G2 F20 SuperTwin Transmission Electron Microscope. Cross-sectional TEM measurements were taken on the JEOL JEM-2100 Transmission Electron Microscope. Cross-sectional QLED samples were prepared using a focused-ion-beam system (Thermo Scientific Scios FIB-SEM). SEM images were obtained using TFS Apreo-S with Lovac Scanning Electron Microscope operating at 2 kV and 13 pA. Current injection-dependent photoluminescence measurements were taken with a Stanford research systems model SR830 DSP lock-in amplifier connected to a Thorlabs optical chopper, driven at 1000Hz, chopping a 405nm laser driven by a CL-2000 diode pumped Crystal Laser power supply. Correlated photoluminescence-electroluminescence imaging was performed on a Photon etc. microscope setup, with a Nikon Intensilight C-HGFI white light source passing through a 450nm short pass filter as the excitation source. Broadband PL emission is collected after a 500nm long pass filter. All XPS spectra were taken on a KratosAxis-Ultra DLD spectrometer. This instrument has a monochromatized Al Kα X-ray and a low energy electron flood gun for charge neutralization. X-ray spot size for these acquisitions was on the order of $700 \times 300$ μm. Pressure in the analytical chamber during spectral acquisition was less than $5 \times 10^{-9}$ Torr. Pass energy for survey and detailed spectra (composition) was 80 eV. Pass energy for the high-resolution spectra was 20 eV. The take-off angle (the angle between the sample normal and the input axis of the energy analyzer) was 0° (0 degree take-off angle ~ 100 Å sampling depth). The KratosVision2 software program was used to determine peak areas and to calculate the elemental compositions from peak areas. CasaXPS was used to peak fit the high-resolution spectra. For the high-resolution spectra, a Shirley background was used, and all binding energies were referenced to the C ls C-C bonds at 285.0 eV. Temperature-dependent PL measurements were performed with the samples on a Linkam LTS420E-P temperature control stage coupled to a Linkam T95-HS temperature controller.

**Author contributions**
G.S., S.R.M, and D.S.G. conceived of the project. G.S. performed the quantum dot syntheses, device fabrication, and device characterization for the project and prepared the manuscript. Y.Z., S.B., and S.R.M. provided key interfacial materials, 2PACz and t-Bu-2PACz, and intellectual discussion for the work. E.R engaged in intellectual discussions. H.C. conducted the scanning electron microscopy (SEM) imaging for the work, while Y.X. conducted the cross-sectional TEM, and J.J. conducted time-resolved photoluminescence (TRPL) measurements. J.K. helped to set up and perform the current injection-dependent photoluminescence setup. C.S. conducted ultraviolet photoelectron spectroscopy (UPS) measurements for the 2PACz and t-Bu-2PACz-modified ITO layers as well as bare ITO and the quantum dot films. All authors contributed to the editing of the manuscript.

**Supporting Information**
Supplementary figures showing synthetic schemes for 2PACz and t-Bu-2PACz, ultraviolet photoelectron spectroscopy results, enlarged cross-sectional TEM images of the QD active layers, TEM images of QDs cast from solution, CIE color coordinates for PVK, 2PACz, and t-Bu-2PACz devices, time-resolved photoluminescence data of QD films and solution, temperature-dependent characterization of films and red-tail fittings of electroluminescence spectra, SEM images of films, dark field images of devices, XPS data, and correlated photoluminescence-electroluminescence microscopy images.

**Acknowledgements**
This work was supported by the National Science Foundation under the STC Grant No. DMR-2019444, for the Center for Integration of Modern Optoelectronic Materials on Demand (IMOD). The authors acknowledge the use of facilities and instruments at the Photonics Research Center (PRC) at the Department of Chemistry, University of Washington, as well as that at the Research Training Testbed (RTT), part of the Washington Clean Energy Testbeds (WCET) system. XPS, TEM, SEM, and profilometry measurements were conducted at the Molecular Analysis Facility, a National Nanotechnology Coordinated Infrastructure (NNCI) site at the University of Washington which is supported in part by the National Science Foundation via awards NNCI-1542101 and NNCI-2025489, the Molecular Engineering & Sciences Institute, and the Clean Energy Institute.

# Supporting Information

# Increased Brightness and Reduced Efficiency Droop in Perovskite Quantum Dot Light-Emitting Diodes using Carbazole-Based Phosphonic Acid Interface Modifiers

Gillian Shen[1], Yadong Zhang[2], Julisa Juarez[1], Hannah Contreras[1], Collin Sindt[3], Yiman Xu[4], Jessica Kline[1], Stephen Barlow[2], Elsa Reichmanis[4], Seth R. Marder[2,3,5], David S. Ginger[1]

[1]Department of Chemistry, University of Washington, Seattle, WA 98195, United States

[2]Renewable & Sustainable Energy Institute, University of Colorado Boulder, Boulder, CO 80309, United States

[3]Department of Chemical and Biological Engineering, University of Colorado Boulder, Boulder, CO 80309, United States

[4]Department of Chemical and Biomolecular Engineering, Lehigh University, Bethlehem, PA 18015, United States

[5]Department of Chemistry, University of Colorado Boulder, Boulder, CO 80309, United States

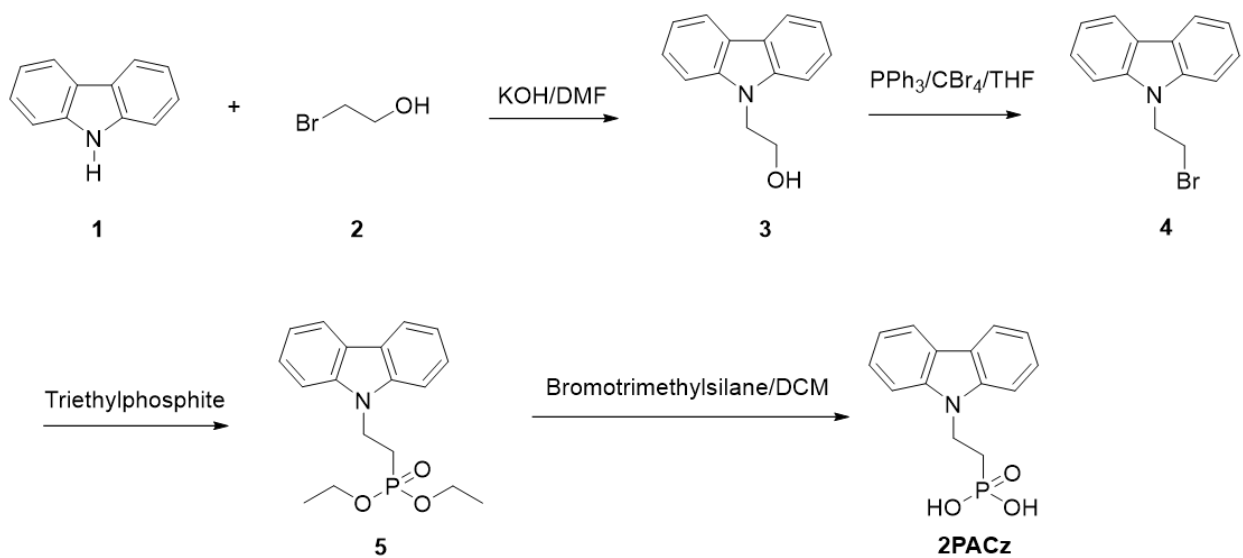

**Figure S1**. Synthesis of [2-(9*H*-carbazol-9-yl)ethyl]phosphonic acid (**2PACz**). $^1$H NMR, $^{13}$C{$^1$H} NMR and $^{31}$P{$^1$H} NMR spectra of diethyl [2-(9*H*-carbazol-9-yl)ethyl]phosphonate (**5**) in CDCl$_3$ are shown in Figure S15. $^1$H NMR, $^{13}$C{$^1$H} NMR and $^{31}$P{$^1$H} NMR spectra of **2PACz** in DMSO-d$_6$ are shown in Figure S16.

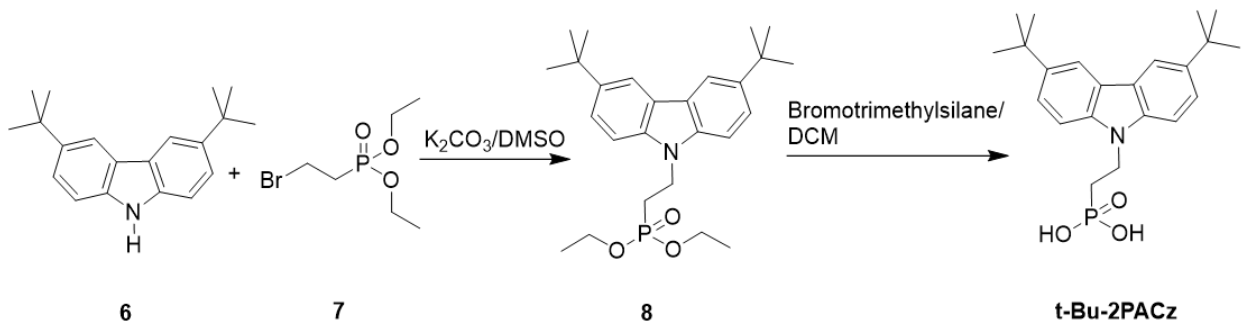

**Figure S2**. Synthesis of [2-(3,6-di(*tert*-butyl)-9*H*-carbazol-9-yl)ethyl]phosphonic acid (**t-Bu-2PACz**). $^1$H NMR, $^{13}$C{$^1$H} NMR and $^{31}$P NMR spectra of diethyl [2-(3,6-di-*tert*-butyl-9*H*-carbazol-9-yl)ethyl]phosphonate (**8**) in CDCl$_3$ are shown in Figure S17. $^1$H NMR, $^{13}$C{$^1$H} NMR and $^{31}$P{$^1$H} NMR spectra of **t-Bu-2PACz** in DMSO-d$_6$ are shown in Figure S18.

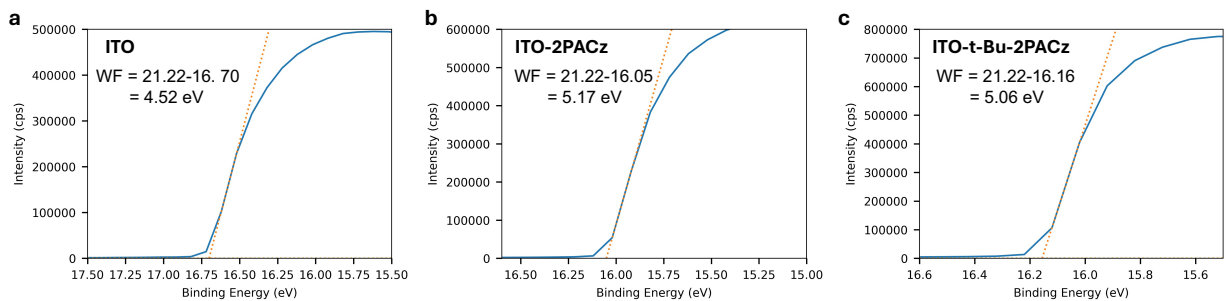

**Figure S3.** Ultraviolet Photoelectron Spectroscopy (UPS) measurements and fittings to determine the work functions of ITO, ITO-2PACz, and ITO-t-Bu-2PACz respectively. a) UPS showing cutoff energy fitting for ITO b) for ITO-2PACz, c) for ITO-t-Bu-2PACz

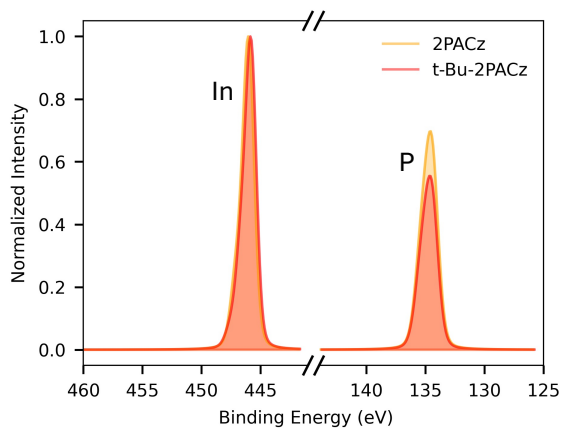

**Figure S4.** XPS detailed scans of In and P Indicating relative surface coverage of 2PACz and t-Bu-2PACz on ITO. Intensities are normalized to indium peak maxima and adjusted for elemental relative sensitivity factors (RSF). $RSF_{In}$ = 7.265, $RSF_P$ = 0.486. We calculate relative elemental ratios from integrated areas under each peak, which reveals a higher molar density of surface coverage of 2PACz on the sample surface, repeated across N=3 samples. These results indicate that the absolute molar coverage of the ITO surface is lower for t-Bu-2PACz, likely due to stearic effects from the tert-butyl group creating a less dense p-p stacking of the carbazole moieties, [5] yet the presence of the aliphatic moiety enables improved wetting of the surface, which we verify with correlated EL-PL measurements (Figure S14). Atomic percentages are tabulated in Table S1.

| Peak | Type | BE (eV) | RSF | 2PACz | | t-Bu-2PACz | |
|---|---|---|---|---|---|---|---|
| | | | | RSF Adjusted Area | Atomic (%) | RSF Adjusted Area | Atomic (%) |
| In | 3d | 446 | 7.265 | 4260.7 | 54.2 | 3767.1 | 60.0 |
| P | 2p | 134.5 | 0.486 | 3606.2 | 45.8 | 2509.8 | 40.0 |

**Table S1.** Calculated atomic percentages of In and P via XPS at the surface of ITO-2PACz and ITO-t-Bu-2PACz respectively.

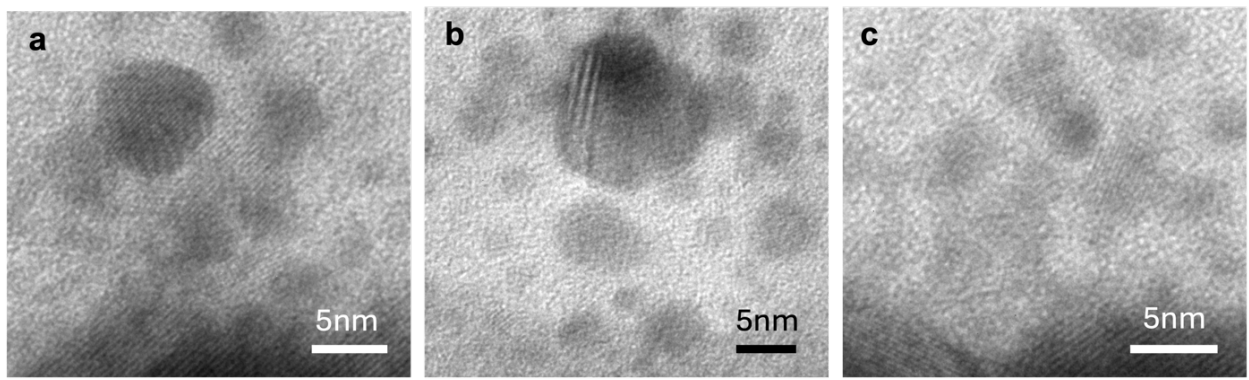

**Figure S5.** Enlarged cross-sectional transmission electron microscopy (TEM) images of active layer of device showing FAPbBr$_3$ Quantum Dots. a-c) cross-sectional TEM images of FAPbBr$_3$ Quantum Dots in a device, showing stochastic variations in particle size and interparticle spacing across different regions of the device.

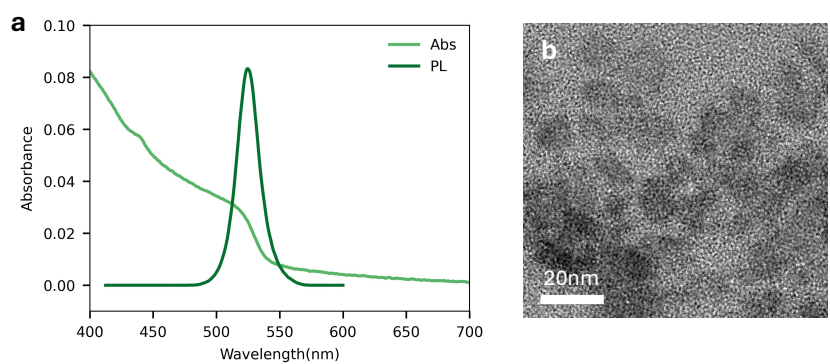

**Figure S6.** QD Solution Characterization. a) Absorbance and PL characteristics of the quantum dots in solution. b) TEM of solution-cast QDs on TEM grid

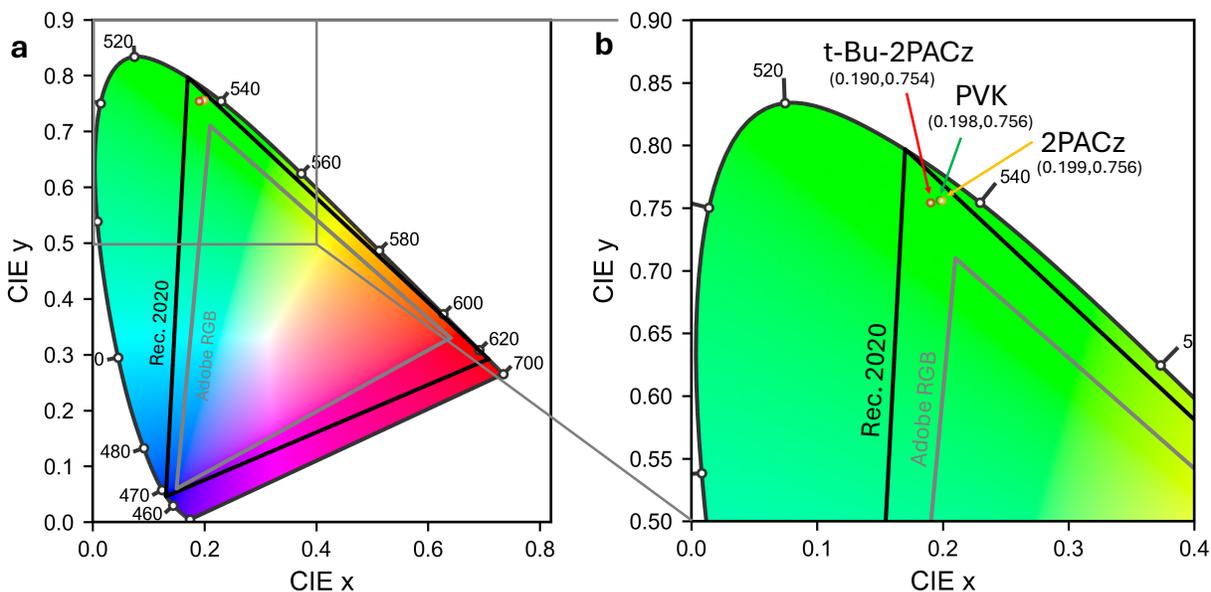

**Figure S7.** Commission Internationale de l'éclairage color coordinates for emission from PVK, 2PACz and t-Bu-2PACz Devices. a) full CIE color chart, with Rec. 2020 and Adobe RGB standards outlined b) enlarged detail of CIE color chart in the green regime, with CIE color coordinates of PVK, 2PACz, and t-Bu-2PACz devices.

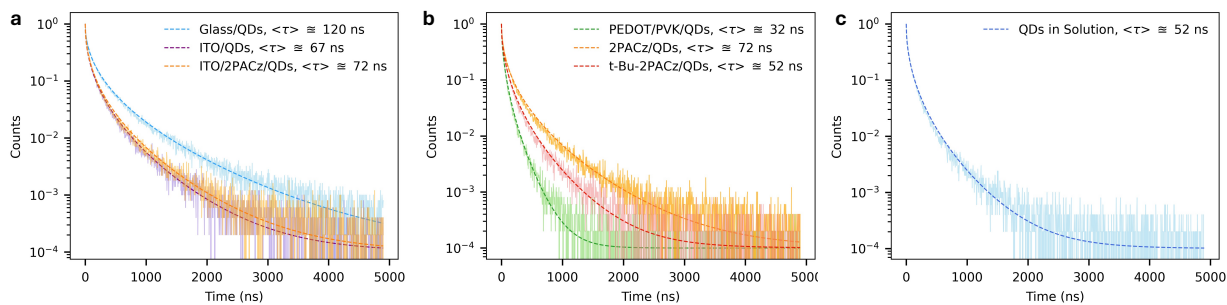

**Figure S8.** Time-resolved photoluminescence (TRPL) data of QDs in solution and in films on various transport interfaces, measured 3 nJ/cm$^2$. a-c) comparison of TRPL traces between QDs a) on glass, ITO, and ITO-2PACz b) PEDOT/PVK, 2PACz, and t-Bu-2PACz, c) in solution. Calculated lifetimes from stretched exponential fittings are detailed in Table S2.

| Sample | $\tau_c$ (ns) | $\langle\tau\rangle$ (ns) | $\beta$ |
|---|---|---|---|
| Glass/QDs | 54 | 120 | 0.47 |
| ITO/QDs | 30 | 67 | 0.47 |
| ITO/PEDOT:PSS/PVK/QDs | 18 | 32 | 0.53 |
| ITO-2PACz/QDs | 32 | 72 | 0.46 |
| ITO-t-Bu-2PACz/QDs | 25 | 52 | 0.49 |
| QDs in Solution | 25 | 52 | 0.49 |

**Table S2.** Calculated characteristic lifetimes $\tau_c$ and average lifetimes $\langle\tau\rangle$ for FAPbBr$_3$ QDs on various substrates and transport layers according to a stretched exponential fitting detailed by Taddei et al. [2]. The fitting assumes a distribution of carrier lifetimes across the film, with $\beta$ values approaching 1 representing a more homogenous lifetime distribution, while $\beta$ values approaching 0 represent a more heterogeneous lifetime distribution.

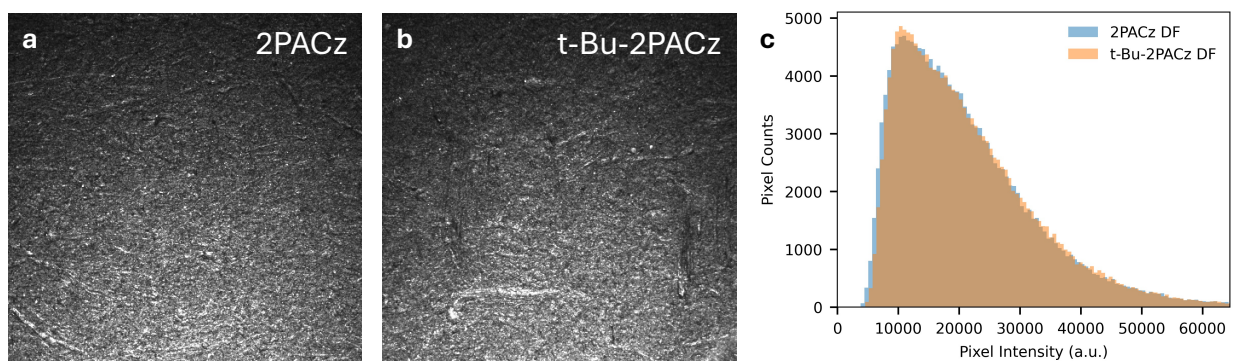

**Figure S9**. Dark Field images of QDs on 2PACz and t-Bu-2PACz a) dark field image of ITO/2PACz/QD half stacks and a) of ITO/t-Bu-2PACz/QD half stacks c) distribution of pixel intensities for each sample, indicating that outcoupling behaviors are largely consistent between the two samples.

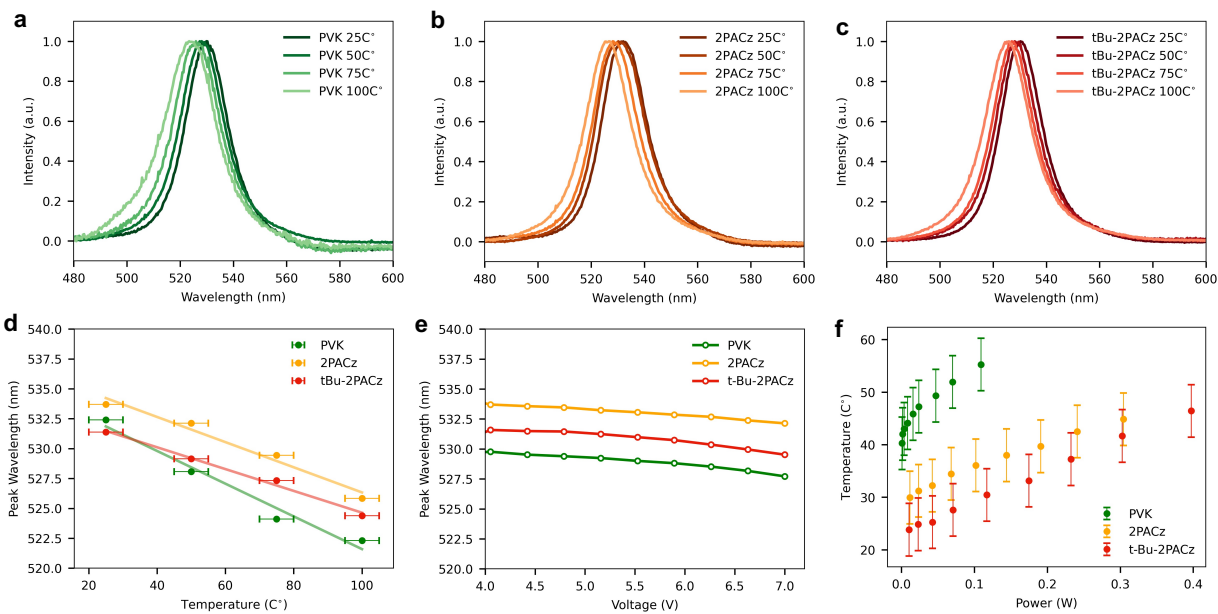

**Figure S10.** Temperature dependent photoluminescence characteristics and extracted local temperature from device electroluminescence. a-c) temperature-dependent photoluminescence characteristics of PEDOT:PSS/PVK, 2PACz, and t-Bu-2PACz devices at 25, 50, 75, and 100 °C d) temperature-dependent peak PL wavelengths of PEDOT:PSS/PVK, 2PACz, and t-Bu-2PACz devices. e) driving voltage-dependent peak emission wavelengths of each device f) power-dependent extracted active layer temperatures from peak wavelengths at each driving voltage and current level.

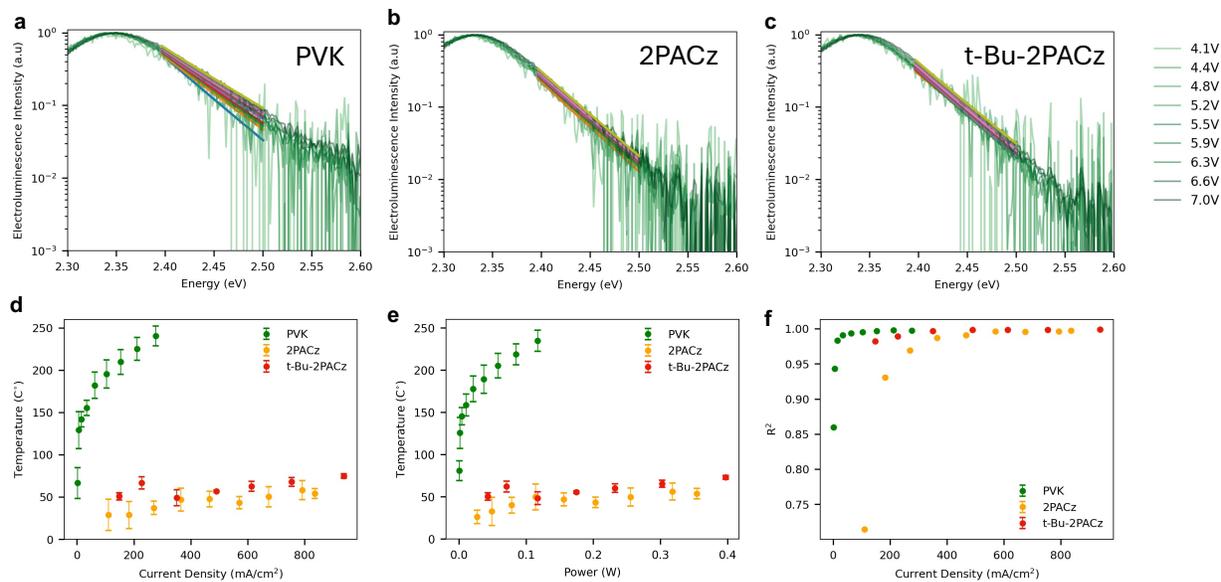

**Figure S11.** Generalized Plank Law fits to electroluminescence profiles of PVK, 2PACz, and t-Bu-2PACz devices. a-c) Fit lines to normalized electroluminescence spectra for PVK, 2PACz, and t-Bu-2PACz devices respectively. Note that signal-to-noise ratio improves at higher drive voltages as electroluminescence intensity increases d) current density-dependent and e) power-dependent emitter temperatures in PVK, 2PACz, and t-Bu-2PACz devices with error bars indicating standard deviations across N=3 devices under each condition. f) $R^2$ values for fits at each current density, indicating improved fits at higher current densities with improved signal-to-noise ratios. We employed the generalized Plank Law equation, otherwise known as the Lasher-Stern-Würfel (LSW) equation ($I_{PL}(E) \propto \frac{E^2}{\exp\left(\frac{E-\Delta\mu}{kT}\right)}$ where E is photon energy, $\Delta\mu$ is the quasi-Fermi level splitting corresponding to the bandgap of the material, k is the Boltzmann constant, and T is the local temperature), [3,4] to perform this fitting. The results reveal the suppression of emissive layer heating at high driving currents in 2PACz and t-Bu-2PACz devices while heating in PVK devices is more significant. A comparison of the fits to the ground truth electroluminescence profiles reveals the tight correspondence between the two (Figure S9), while $R^2$ values for the fits improve to >0.99 at higher driving currents, as the signal-to-noise ratio improves in the electroluminescence spectral profile.

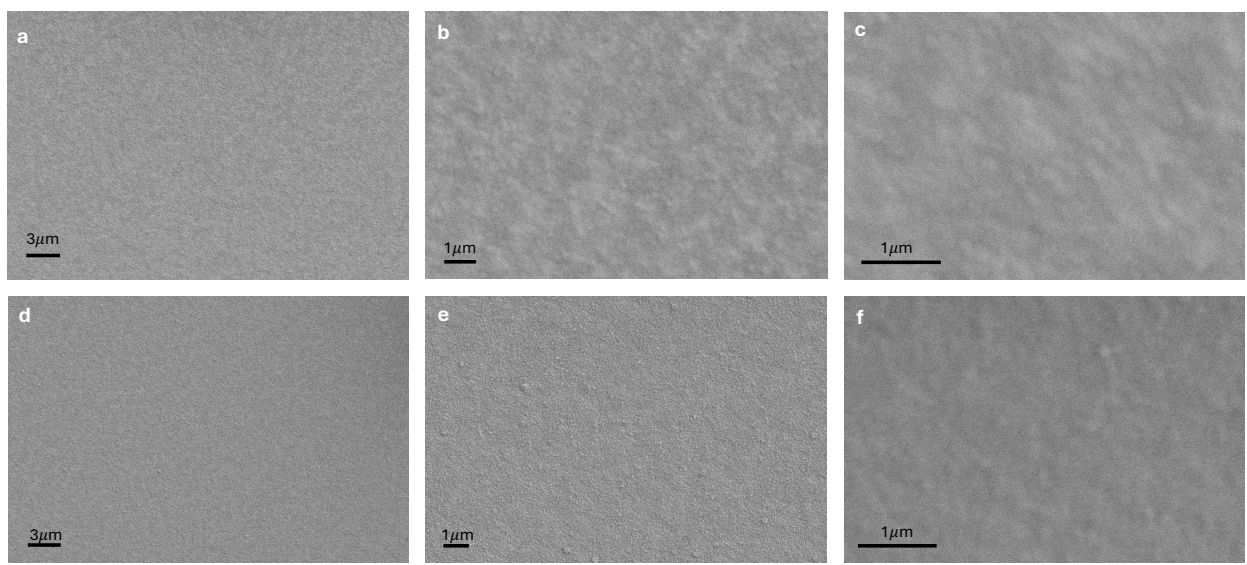

**Figure S12.** Scanning Electron Microscopy (SEM) images of half stacks with QDs a)-c) ITO/2PACz/QDs d)-f) ITO/t-Bu-2PACz/QDs at various levels of magnification.

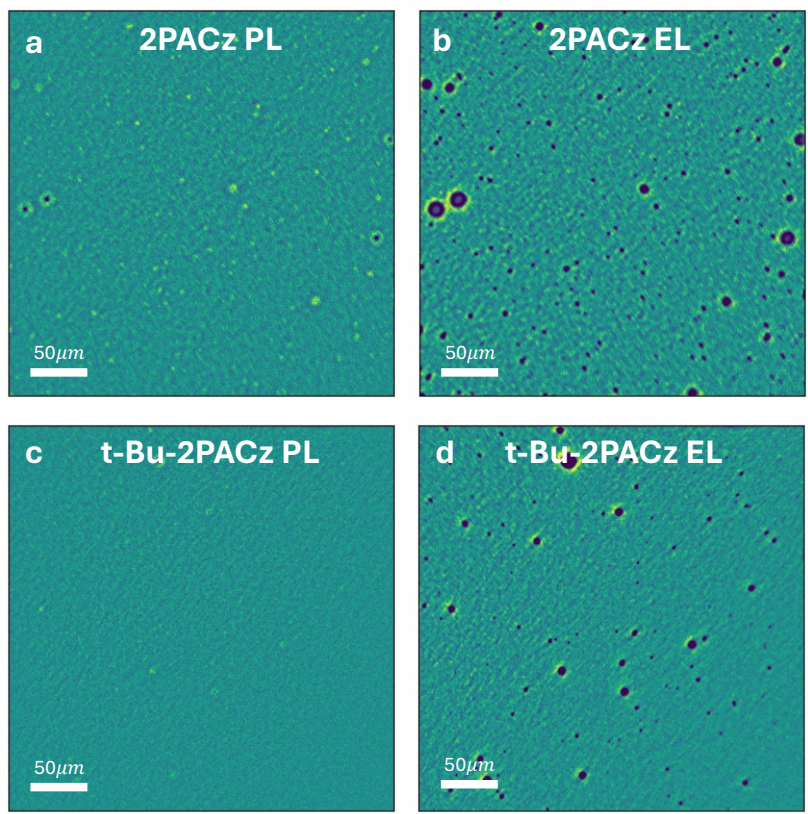

**Figure S13.** Correlated photoluminescence-electroluminescence microscopy images of devices. a) 2PACz photoluminescence b) 2PACz electroluminescence c) t-Bu-2PACz photoluminescence d) t-Bu-2PACz electroluminescence.

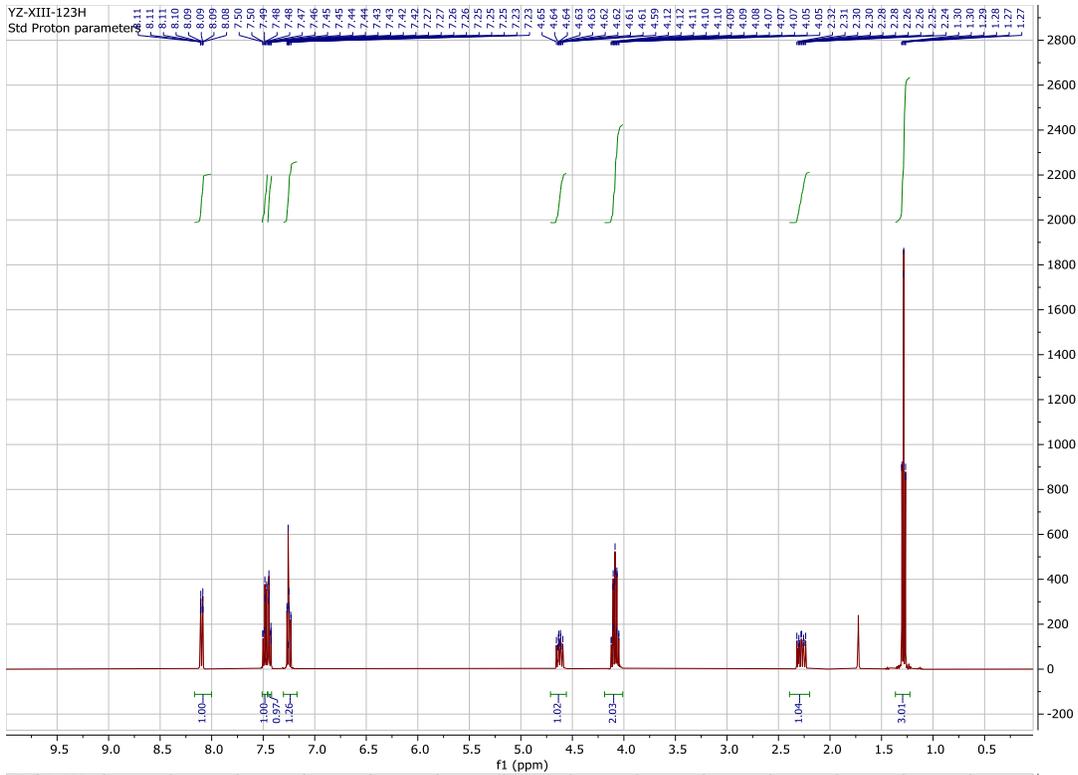
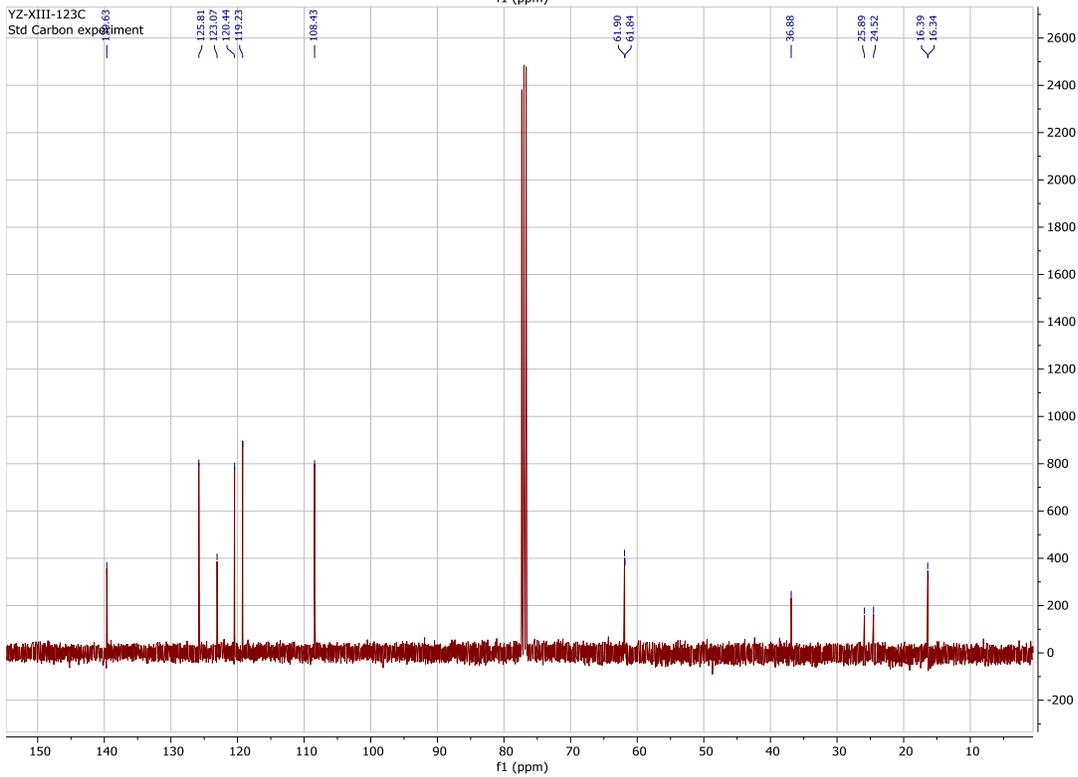

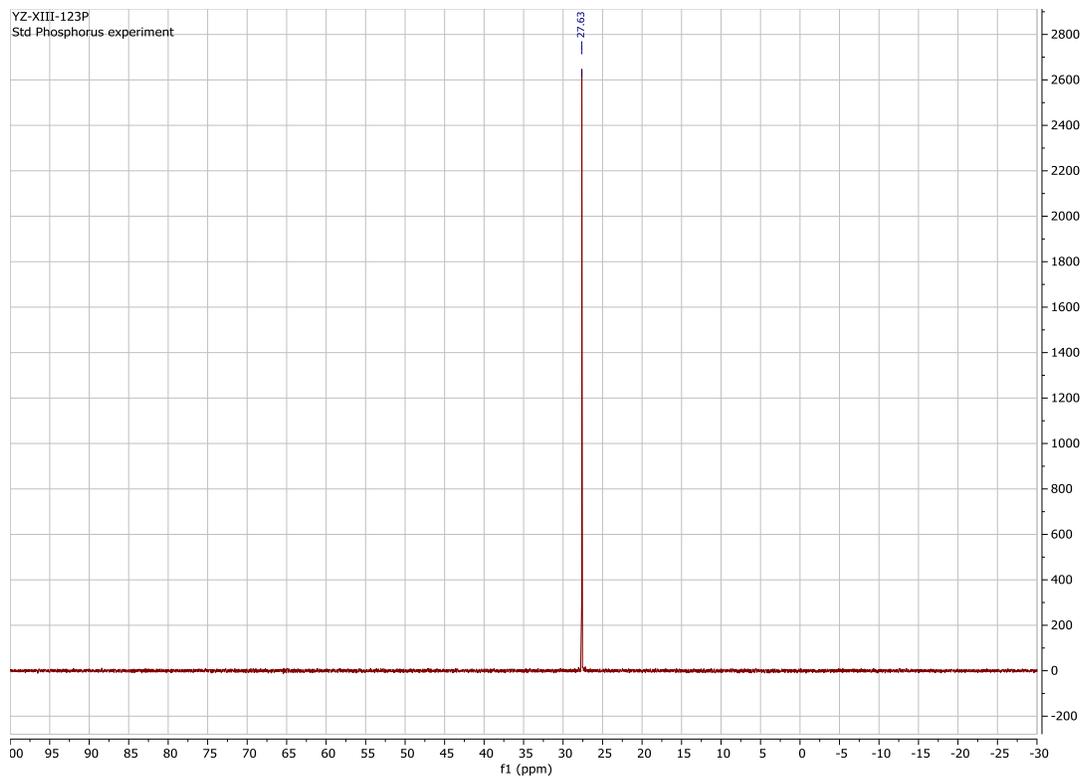

**Figure S14.** ¹H NMR, ¹³C{¹H} NMR and ³¹P{¹H} NMR spectra of diethyl [2-(9*H*-carbazol-9-yl)ethyl]phosphonate (**5**) in CDCl₃.

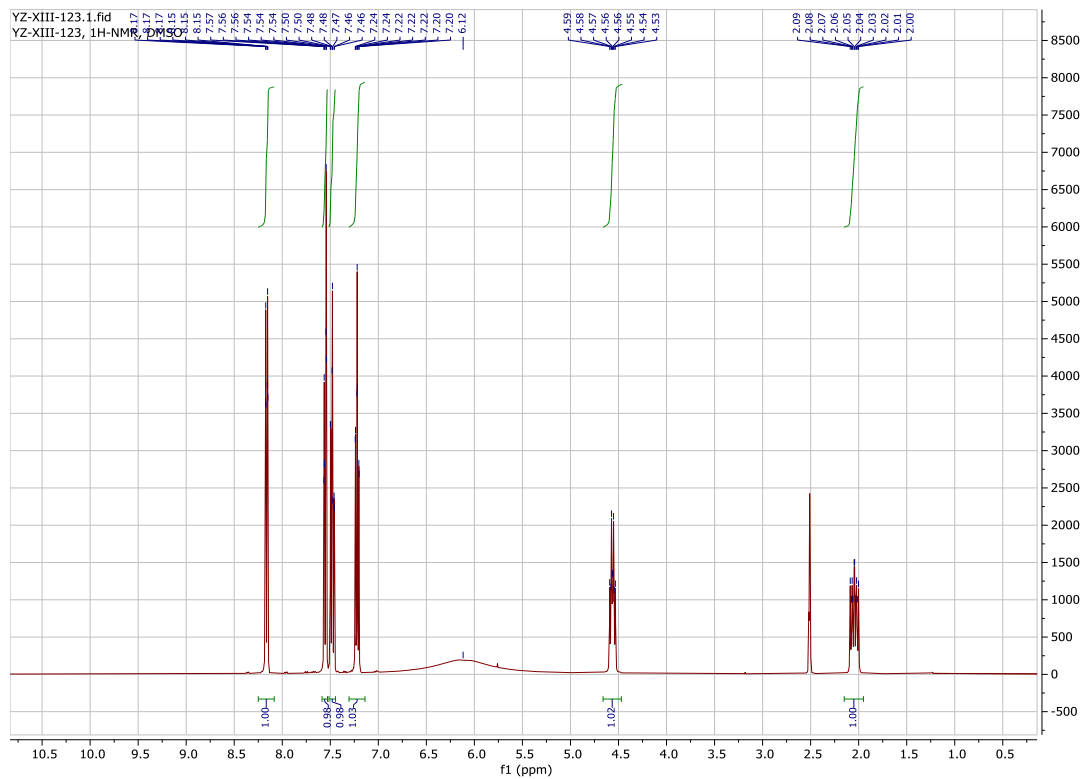

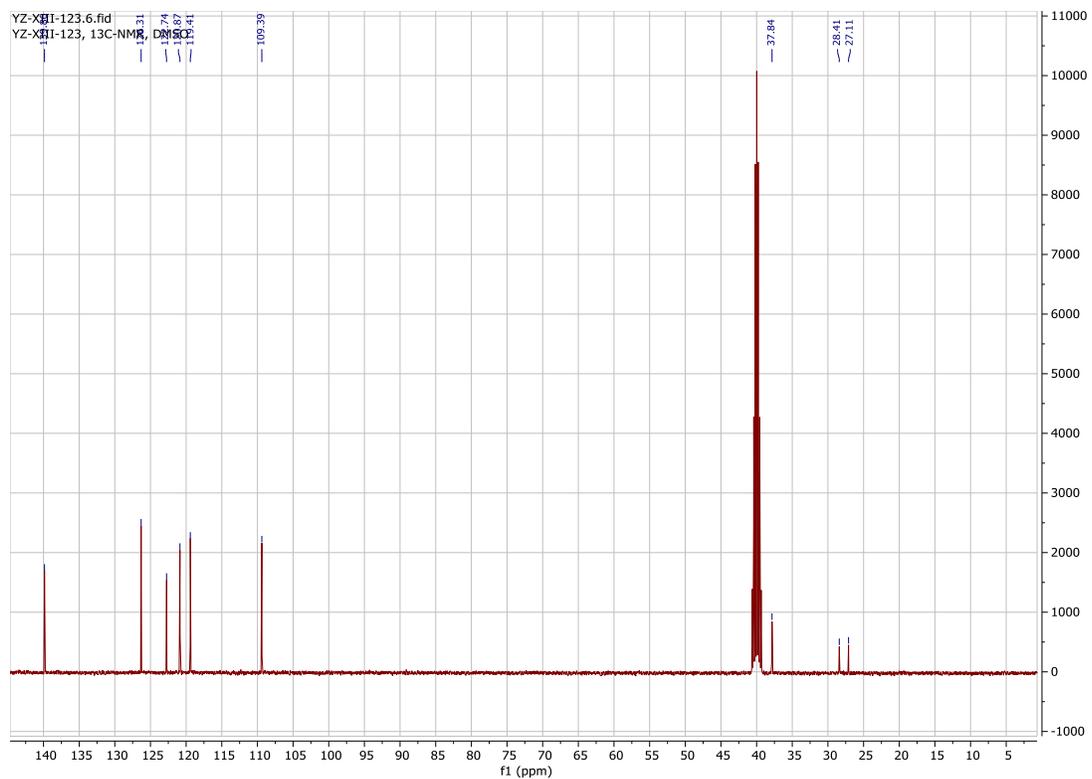
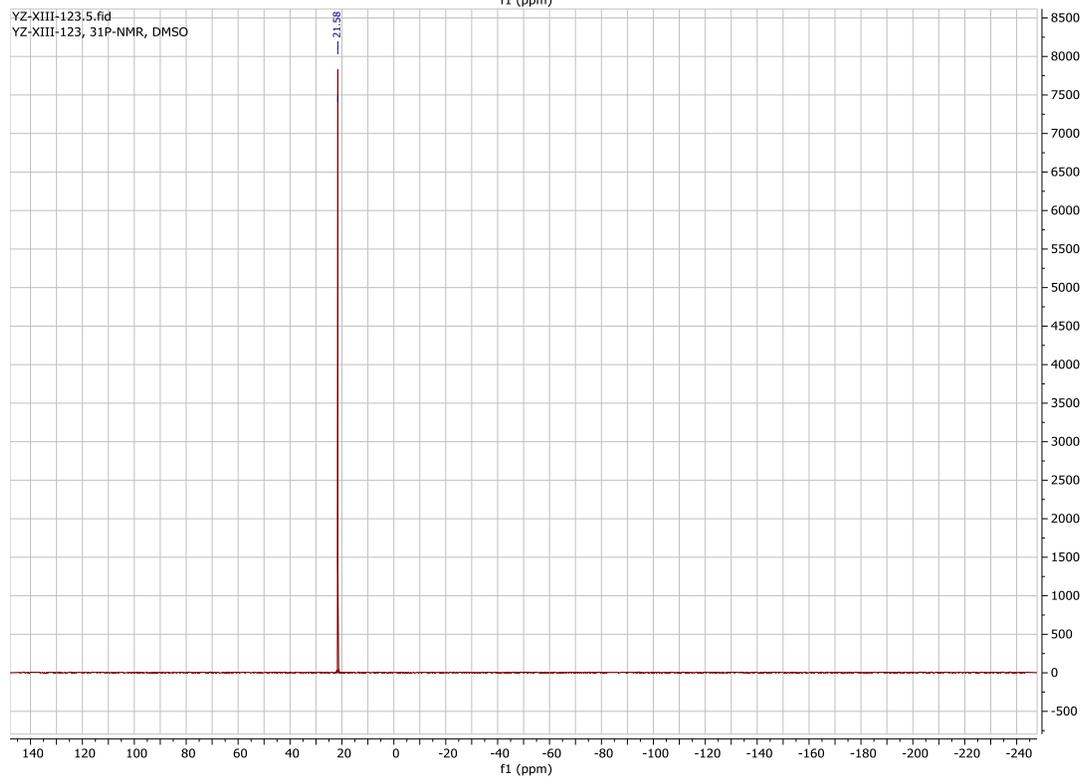

**Figure S15.** $^1$H NMR, $^{13}$C{$^1$H} NMR and $^{31}$P{$^1$H} NMR spectra of **2PACz** in DMSO-d$_6$.

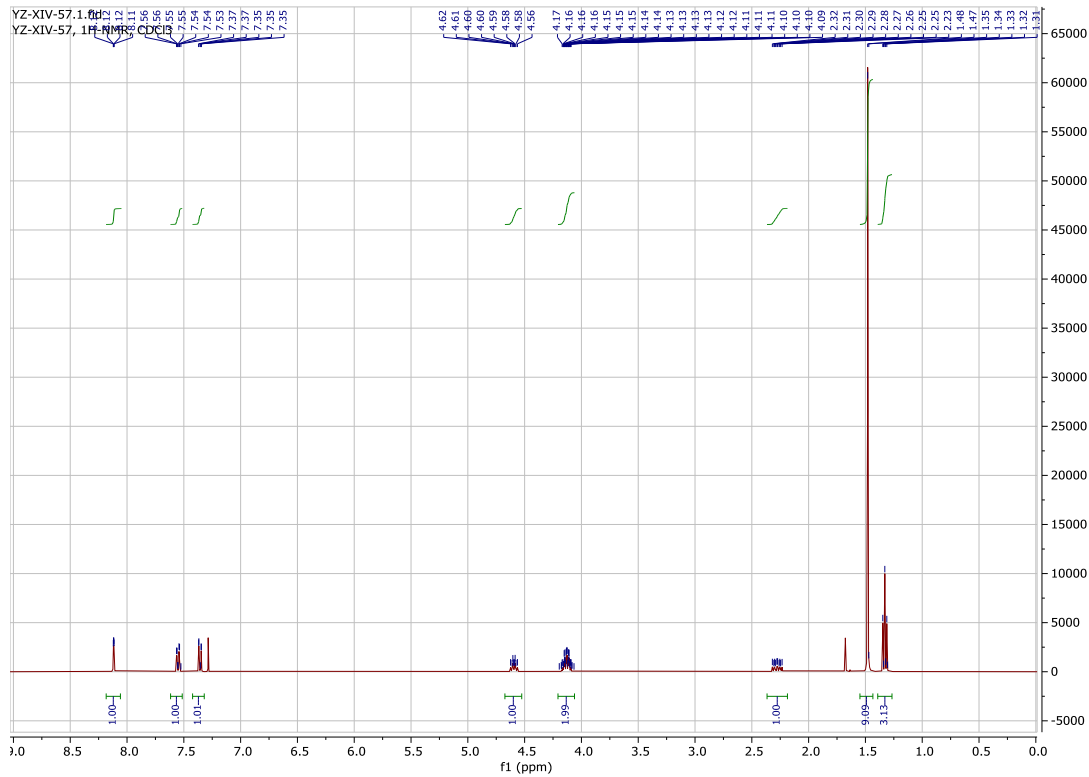
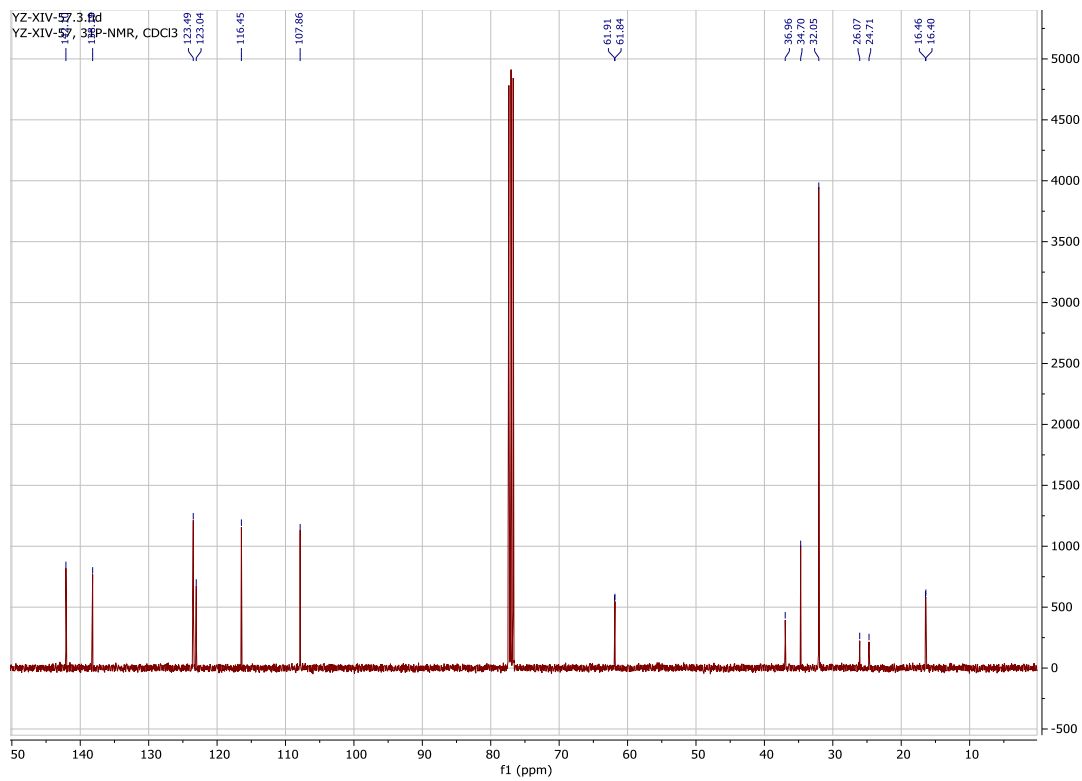

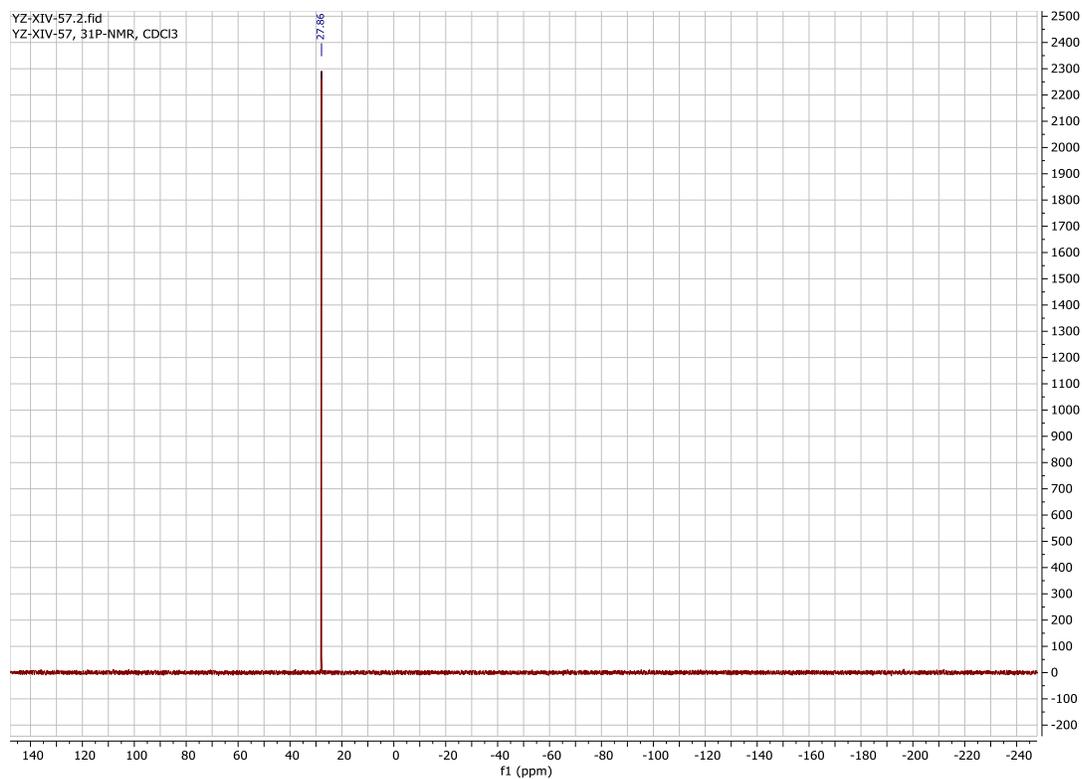

**Figure S16.** ¹H NMR, ¹³C{¹H} NMR and ³¹P NMR spectra of diethyl [2-(3,6-di-*tert*-butyl-9*H*-carbazol-9-yl)ethyl]phosphonate (**8**) in CDCl₃.

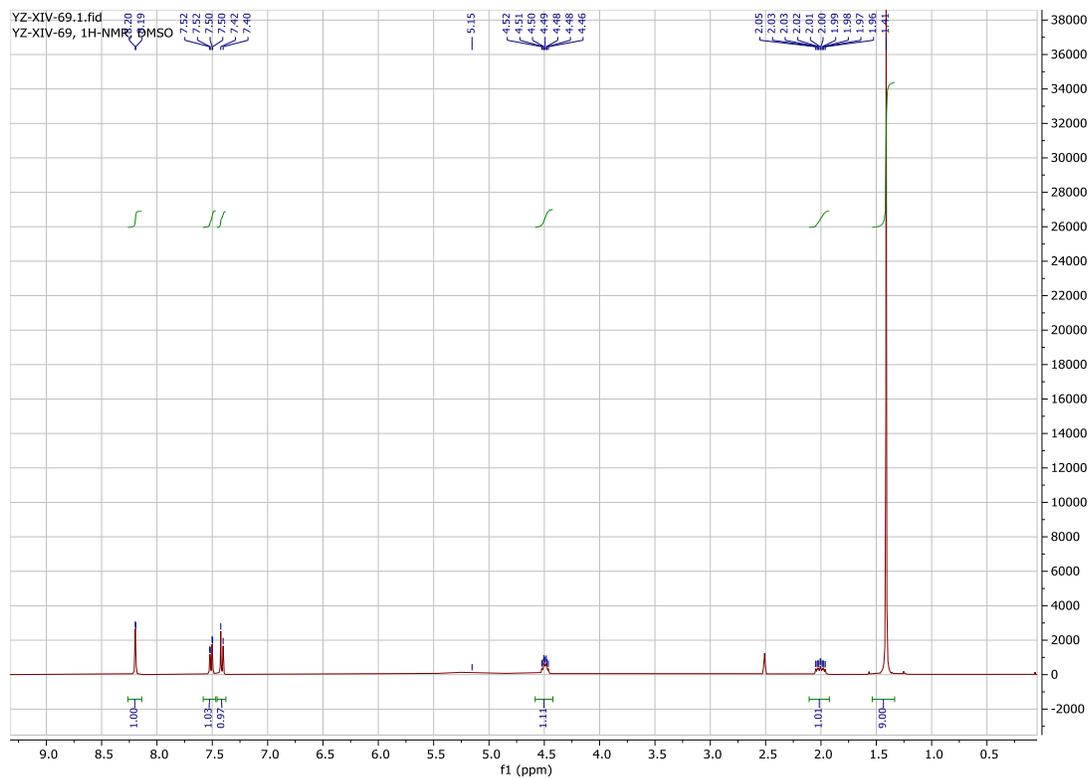

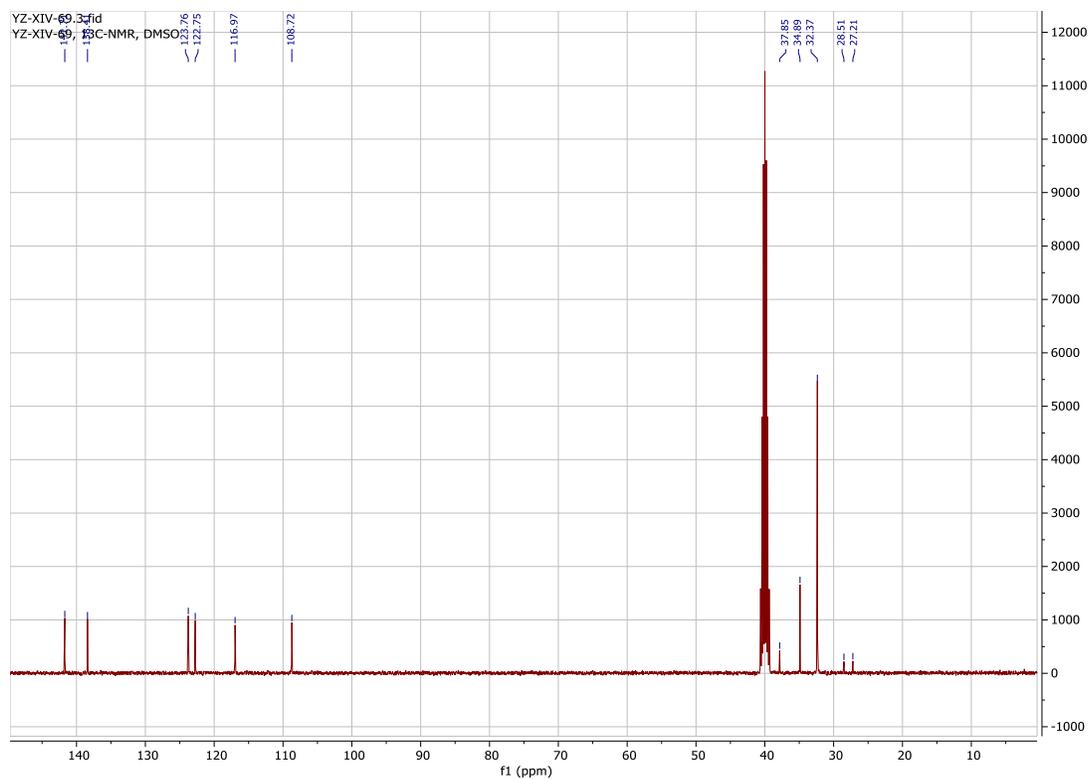
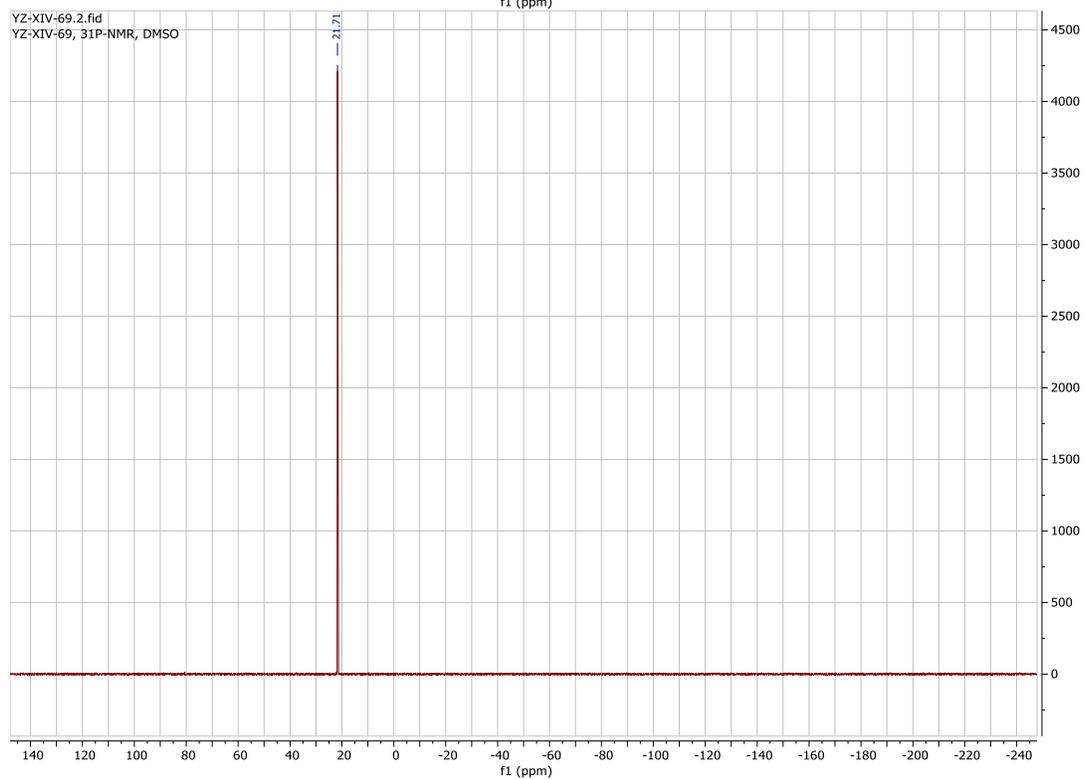

**Figure S17.** $^1$H NMR, $^{13}$C{$^1$H} NMR and $^{31}$P{$^1$H} NMR spectra of **t-Bu-2PACz** in DMSO-d$_6$.